# A Semantic Geometry for Uncovering Paradigm Dynamics via Scientific Publications

Jinchang Liu, Qingshan Zhou*, Hongkan Chen*, and Yi Bu*

*Department of Information Management, Peking University, Beijing 100871, China*

*Correspondence: Qingshan Zhou (zqs@pku.edu.cn), Hongkan Chen (chenhongkan@pku.edu.cn), and Yi Bu (buyi@pku.edu.cn).

**Abstract**: Science advances not only by accumulating discovered patterns but by changing how new problems and solutions are expressed. While structural indicators track scholarly attention, they offer only an indirect proxy for the reorganization of meaning. We propose a semantic geometry based on the R–P–C (references, focal publication, and citing publications) framework to quantify how a publication positions itself relative to its knowledge base and diffusion. This geometry identifies three publication types: consolidating, exploratory and balanced. Our results show that the semantic similarity and distance between a publication's knowledge base and diffusion serve as a semantic foundation for disruption, with novelty (atypical reference combinations) acting as an antecedent disturbance that triggers a semantic rupture. This is related to team size, where small teams preserve a higher potential for exploratory departures while large collaborations systematically align with paradigmatic consolidation. Crucially, this geometry explains why citation trajectories differ; consolidating research earns rapid recognition by lowering comprehension costs, while exploratory work faces high paradigm conversion costs that result in slower, more selective diffusion. Collectively, this R–P–C framework provides a robust instrument for monitoring the dynamics of scientific paradigms.

**Keywords:** paradigm dynamics; semantic relatedness; disruption; citation impact; scientometrics.

## 1 Introduction

Science advances entail both the accumulation of findings and the ongoing reorganization of scientific language (Fortunato, Bergstrom, Börner, Evans, Helbing, Milojević, Petersen, Radicchi, Sinatra, & Uzzi, 2018; Milojević, 2015). As Thomas Kuhn argued, the development of science is punctuated by paradigm shifts: when established frameworks fail to explain an expanding set of anomalies, confidence in the prevailing paradigm erodes, a period of crisis ensues, and a new paradigm gradually emerges to replace the old (Kuhn, 1962). Such transformations do not simply mean changes in data or findings; they are accompanied by redefinitions of problems, the adoption of new methods and instruments, and, most visibly, the transformation of scientific language. New problems and methods inevitably require new terminology to describe them. In this sense, paradigm shift is registered both cognitively and semantically: it reshapes not only what scientists study but also how they speak about it.

To capture these dynamics, quantitative research on scientific change has predominantly relied on structural indicators derived from the topology of citation network. Prominent indicators, e.g., citation count, the disruption index (Wu et al., 2019), and novelty score (Luo et al., 2022), track how a focal publication connects to its predecessors and successors. These indicators rest on the assumption that the topology of citation network mirrors the conceptual organization of a field (Smith, 1981). However, they offer only an indirect proxy. While they capture the flow of attention and credit, they do not inherently reveal the reorganization of scientific language. Citation practices are often influenced by social conventions, such as strategic referencing or deference to authority (H. Small, 1998), which further complicates the correspondence between citation structure and paradigm shift. In other words, structural indicators measure the channels of communication, leaving the semantic dimension of paradigm shifts largely unmeasured.

To directly trace these paradigm dynamics, recent advances in computational linguistics have begun to conceptualize and measure scientific changes within high-dimensional semantic spaces (Lee & Su, 2010; Li et al., 2016; Lozano et al., 2019; Radhakrishnan et al., 2017; Su & Lee, 2010). By embedding texts in a shared continuous and high-dimensional semantic space, researchers can trace how terminologies cluster, diverge, or reconnect, and how focal publications align with or diverge from their knowledge base (e.g., Mikolov et al., 2013). Within this space, semantically related publications tend to form clusters that represent coherent conceptual areas or problem framings. Shifts in the boundaries or centers of these clusters signal how paradigms consolidate or fragment over time. This semantic perspective provides a parallel dimension through which paradigm dynamics can be observed as gradual reorganizations of knowledge, the evolving alignment between a field's knowledge base and its knowledge diffusion. For example, shifts in the semantic proximity among key terms reveal how emerging concepts diverge from established terminology or how new problem framings displace previously central ones, patterns that may remain invisible in citation networks (A. J. Yang, 2025).

In what follows, we use the semantic dimension of science to trace how paradigms consolidate and reorient. By examining how scientific language shifts across a focal publication, its references, and its citing publications, we aim to capture the reorganization of knowledge at a finer granularity than pure citation structures allow. This approach links semantic change to paradigm dynamics, offering a nuanced picture of the processes through which scientific paradigms shift. Crucially, our unit of analysis captures localized semantic reconfigurations within specific subfields or publication-centered neighborhoods (i.e., micro-paradigms), rather than monolithic, discipline-wide Kuhnian revolutions.

# 2 Related Work

## 2.1 Networked approaches

The quantitative studies of science have been deeply influenced by the concept of paradigm shifts. These shifts involve radical changes in a field's problems, methods, and terminology (Kuhn, 1962). While Kuhn's analysis was historical and descriptive, a central goal for quantitative studies of science has been developing empirical methods to capture these dynamics and to further identify how semantic clusters consolidate or fragment over time.

The dominant and longest-running approach to this challenge uses the structure of citation networks as an indirect proxy for a field's conceptual structure. The assumption is that the architecture of scholarly communication mirrors the (re)organization of knowledge. Classic methods, particularly co-citation analysis (H. G. Small, 1973; White & Griffith, 1981), map a field's intellectual structure by positing that frequently co-cited papers represent coherent conceptual areas. Consequently, a paradigm shift is identified through the emergence of distinct new clusters or a decisive shift in the center of research attention. Other approaches, like main path analysis (Hummon & Dereian, 1989), trace a field's developmental trajectories through the citation network to identify the dominant lines of research, where new paradigms may emerge as divergent paths.

More recently, specific indicators have been proposed to capture the role of individual publications in paradigm shifts. These include the novelty index, which identifies atypical combinations of references (Luo et al., 2022; Uzzi et al., 2013), and the disruption index, which measures how a publication structurally displaces its predecessors in the citation record (Funk & Owen-Smith, 2017; Wu et al., 2019). Recent advancements innovatively utilize citation graph embeddings to create a robust, continuous measure of this structural disruption (M. Kim et al., 2026). While their approach focuses on how transformative works reconfigure trajectories within the citation network structure, our R-P-C framework focuses on the reorganization of scientific language. Together, these structural and semantic approaches provide highly complementary signals, capturing both the shifts in scholarly attention and the conceptual divergence of knowledge content.

## 2.2 Semantic approaches

The second branch of approaches seeks to complement network-based indicators by directly analyzing the scientific language of publications. This semantic perspective (Lee & Su, 2010; Li et al., 2016) leverages natural language processing techniques to provide a direct view onto how problem framings and terminology evolve.

Early efforts in this area often relied on topic modeling to identify latent thematic structures in a large corpus and track changes in their popularity over time (Blei, 2012; Griffiths & Steyvers, 2004). A paradigm shift, in this view, could be seen as the decline of established topics and the emergence of new ones (Hall et al., 2008). More recent advances utilize high-dimensional text embeddings (Cohan et al., 2020) to capture nuanced semantic relatedness. This has enabled the measurement of cognitive proximity (H. Kim et al., 2022), the mapping of knowledge diffusion (Chen et al., 2025; M. C. Kim et al., 2020), and the assessment of semantic novelty (Shibayama et al., 2021; Yin et al., 2023).

This semantic research, however, has often focused on isolated pairwise relationships. For example, studies measure the semantic distance between a publication and its references to quantify novelty (Shibayama et al., 2021), or between a publication and its citing publications to track diffusion (M. C. Kim et al., 2020). However, analyzing these relations in isolation fails to capture the continuity of changes. A departure from the past signals novelty, but it does not indicate whether the field followed this new direction. By integrating these dimensions, our unified geometry (see Section 3) captures the complete trajectory, enabling us to distinguish between transient novelty and successful paradigm reorientation.

## 3 Methodology

### 3.1 Data collection and preprocessing

We utilize bibliographic records from OpenAlex, a freely accessible scholarly database with broad coverage of publications, authors, institutions, and citation links across disciplines (Priem et al., 2022). OpenAlex employs a hierarchical "Concept" classification system (Level 0–Level 5) to categorize works based on their textual content. In our analysis, we focus on *Astronomy*—a Level 1 concept under *Physics* (Level 0). As a domain featuring prominently in historical accounts of paradigm shifts—such as the Copernican Revolution (Kuhn, 1962)—it offers a classic context for examining the dynamics of conceptual change; the selection of astronomy in this study also follows the tradition of many science of science prior works (e.g., Milojević, 2014). For consistency and comparability, we restrict works to journal articles and conference publications in our further analyses. To ensure sufficient context for semantic measurement and minimize statistical noise in sparse citation networks, we retain only publications that (i) cite at least ten references (Uzzi et al., 2013), and (ii) have received at least five citations. Applying these criteria yields a final dataset of 668,979 astronomy publications. Sensitivity analyses across varying reference and citation thresholds confirm the robustness of our classification proportions and core findings (see Supplementary Note 4 and Table S1). The distributions of citation counts and reference counts for these publications are reported in Figure S1.

We also use 768-dimensional text embeddings generated by SPECTER, a SciBERT-

based transformer further trained with citation-informed supervision (Cohan et al., 2020). Following standard protocols, these embeddings are generated exclusively from titles and abstracts to ensure standardized text density. While SPECTER effectively captures linguistic context, we acknowledge the potential risk of citation-signal leakage due to its citation-informed supervision. Specifically, such leakage could theoretically confound the interpretation of our semantic measurements by artificially inflating their correlation with structural outcomes like disruption and citations. Furthermore, to rule out this possibility and ensure our framework measures semantics independently of citation structures, we verified that our framework remains completely robust when evaluated using purely text-trained SciBERT embeddings (see Figure S2). Following the original design of this model, we utilized the titles and abstracts provided by OpenAlex as the standard input. Unlike general-purpose embeddings, SPECTER captures both the linguistic content of scientific publications and their citation contexts, making it well-suited for quantifying semantic similarity among scientific publications. Beyond citation links, these embeddings allow us to quantify the semantic proximity and divergence of publications, providing a complementary lens on paradigm dynamics; extensive robustness checks further confirm the stability of this framework (see Supplementary Note 1). For instance, in astronomy, text embeddings reveal how research on exoplanets gradually redefined its problem space—shifting from the language of stellar astrophysics toward that of planetary formation and habitability—a transformation largely masked in citation networks by continued references to foundational stellar studies.

## 3.2 Semantic measurements

Recent work in bibliometrics and science of science has increasingly turned to semantic relatedness as a complement to citation-based measurements (Lee & Su, 2010; Li et al., 2016; Lozano et al., 2019; Radhakrishnan et al., 2017; Su & Lee, 2010). Following prior research, we focus on the semantic relatedness among a focal publication (P), the publication(s) P cites (R), and the publication(s) that cite P (C), hereafter referred to as the R–P–C geometry. This geometry captures how a focal publication bridges its knowledge base and diffusion, thus providing a direct lens on paradigm dynamics: whether a publication remains anchored in established scientific language (R-P), reorients subsequent research (P-C), or reconfigures the relatedness between inherited and emerging clusters (R-C). Specifically, we consider three pairwise relations:

- **P–R (focal publication–references)** captures the degree of novelty relative to the knowledge base, i.e., whether it follows established trajectories or explores new directions.
- **P–C (focal publication–citing publications)** captures how subsequent publications “extend” relative to the focal publication, i.e., whether later research consolidates its framing or branches into new directions.
- **R–C (references–citing publications)** captures the continuity (or “rupture”) between the knowledge base and diffusion, i.e., whether subsequent citing publications remain aligned with a similar semantic direction as the references or shift toward a different semantic cluster, regardless of the focal publication's own semantic position.

We employ both Euclidean distance and cosine similarity while they capture distinct yet complementary aspects of semantic change. Euclidean distance provides an absolute perspective, quantifying the straight-line separation between R–P–C components to measure the magnitude of difference in semantic space (**point-based perspective**). In contrast, cosine similarity offers a proportional perspective, capturing directional divergence regardless of absolute magnitude (**direction-based perspective**), effectively identifying whether a focal publication maintains or departs from the shared orientation of its knowledge base.

**(i) Point-based perspective: Euclidean distance**

When working with Euclidean distance, we treat embeddings as points in a high-dimensional space. For the focal publication $p$, $\mathcal{R}$ and $\mathcal{C}$ denote the sets of references and subsequent citing publications of $p$; $r \in \mathcal{R}$ and $c \in \mathcal{C}$. Let $x_p \in \mathbb{R}^{768}$ be the embedding of $p$, $x_r \in \mathbb{R}^{768}$ and $x_c \in \mathbb{R}^{768}$ the embedding of $r$ and $c$, respectively. We then represent the semantic positions of the focal publication ($p$), its references ($\mathcal{R}$), and its citing publications ($\mathcal{C}$) by their corresponding centroids, $P$, $R$, and $C$:

$$P = x_p \tag{1}$$

$$R = \frac{1}{|\mathcal{R}|}\sum_{r\in\mathcal{R}} x_r \tag{2}$$

$$C = \frac{1}{|\mathcal{C}|}\sum_{c\in\mathcal{C}} x_c \tag{3}$$

While simple averaging provides an intuitive, parameter-free baseline, our paradigm role classifications remain remarkably stable across alternative semantic aggregations, including time-weighted means, citation-weighted means, geometric median, medoids, and PCA-ABTT (see Supplementary Note 1 and Figure S3). These alternative aggregations confirm that our geometric measurements are highly robust against the high semantic dispersion inherent in novel publications (see Supplementary Note 5).

Semantic distances such as $d_{RP}$, $d_{PC}$, and $d_{RC}$ are then defined as the pairwise Euclidean distances among $P$, $R$, and $C$. Formally, the three pairwise distances ($d_{RP}$, $d_{PC}$, and $d_{RC}$) can be written as:

$$d_{ab} = \| a - b \|_2, a, b \in \{R, P, C\}, a \neq b \tag{4}$$

where $|| \cdot | \quad |$ denotes the Euclidean (L2) norm. These three distances form the edges of a semantic triangle with vertices $P$, $R$, and $C$. We further compute the area of this triangle, denoted by $S$, using Heron's formula:

$$S = \sqrt{m(m - d_{RP})(m - d_{PC})(m - d_{RC})} \quad (5)$$

where $m$ denotes the semi-perimeter of the triangle, defined as $m = \frac{1}{2}(d_{RP} + d_{PC} + d_{RC})$.

**(ii) Direction-based perspective: cosine similarity**

When using cosine similarity, we interpret embeddings as vectors. As before, for each focal publication, the reference and citation sets are represented by their centroids, but here $R$ and $C$ are treated as representative semantic orientations of these sets rather than positions. Pairwise similarities are then calculated as:

$$s_{ab} = \frac{a \cdot b}{\| a \|_2 \| b \|_2}, a, b \in \{\overrightarrow{OR}, \overrightarrow{OP}, \overrightarrow{OC}\}, a \neq b \quad (6)$$

where $O$ denotes the origin in the high-dimensional semantic space.

Each of the three relations captures a distinct semantic function regarding orientation. The measure $s_{RP}$ reflects semantic alignment with the knowledge base, indicating whether the publication parallels established trajectories. The measurement $s_{PC}$ captures the directional consistency of subsequent publications relative to the focal publication. By contrast, $s_{RC}$ plays a central role: it captures the conservation of paradigm orientation between the knowledge base ($\mathcal{R}$) and knowledge diffusion ($\mathcal{C}$). Unlike $s_{RP}$ or $s_{PC}$, which only describes how a focal publication positions itself relative to its immediate neighbors in the citation network, $s_{RC}$ reflects the structural reconfiguration of the knowledge flow. A high $s_{RC}$ indicates that the citing community remains anchored to the cluster of references, reinforcing established paradigms; conversely, a low $s_{RC}$ signals a directional rupture, implying that the focal publication mediated a turn toward a distinct semantic cluster.

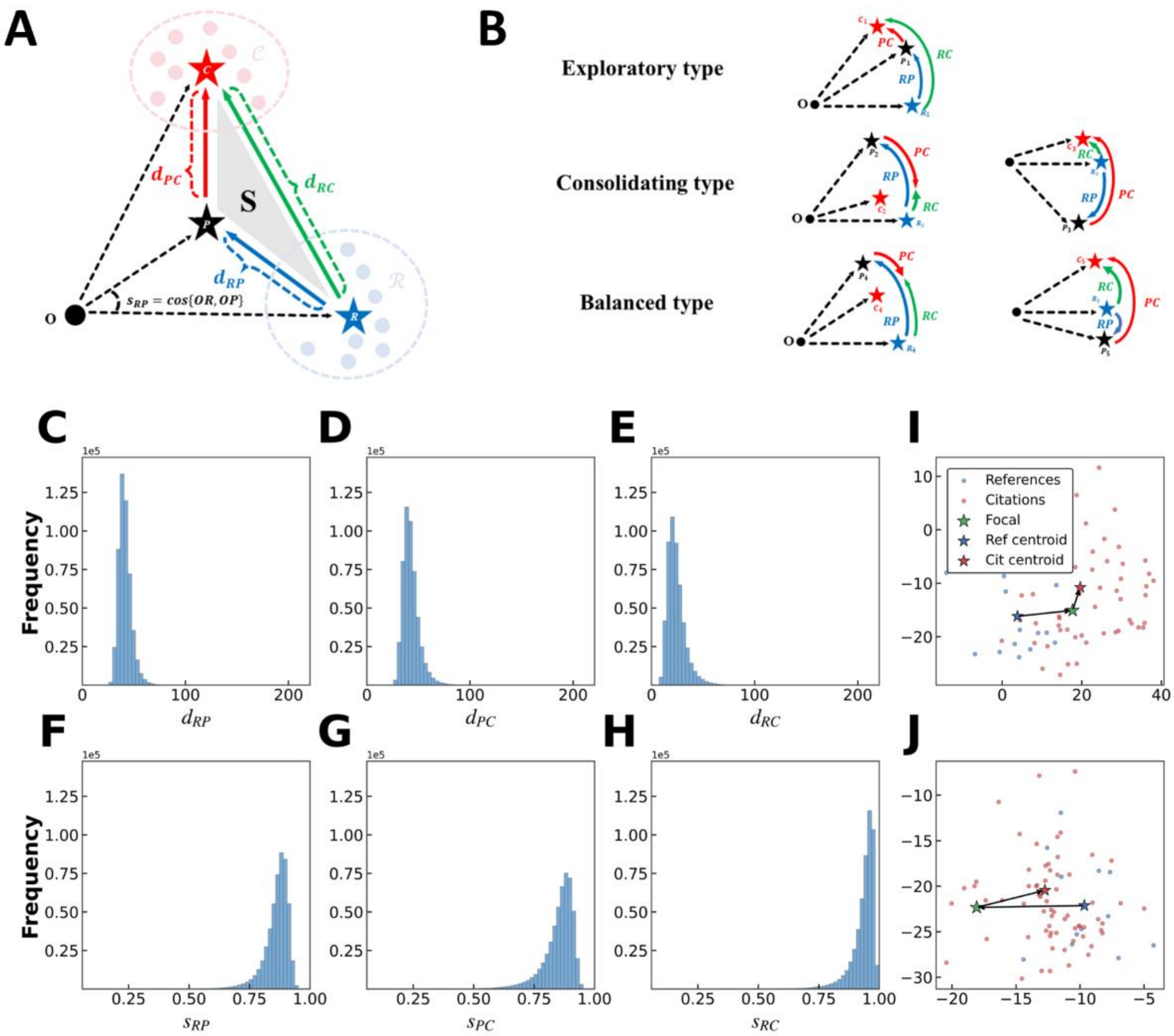

**Figure 1. Semantic geometry of references–publications–citing publications (R–P–C): illustration, exemplars, and disruption patterns.** (A) Schematic of the R–P–C triangle in two-dimensional projection. The black star represents a focal publication. The blue and red stars denote the centroids of the reference (all blue, clustered circles) and citation sets (all red, clustered circles), respectively. The blue arrow indicates the semantic relatedness from references to the focal publication ($R \rightarrow P$), the red arrows represent the relation from the focal publication to its citing publications ($P \rightarrow C$), and the green arrow captures the direct relation between references and citing publications ($P \rightarrow C$). Dotted lines represent embeddings (marked as "straight lines" between $R/P/C$ to $O$, the origin of the high-dimensional semantic space). (B) Classification logic based on the relative magnitudes of $s_{RP}$, $s_{PC}$, and $s_{RC}$. (C–H) Distributions of similarity and distance measurements across all publications. (I–J[1]) Representative cases: *Fundamentals of XAFS* (Newville, 2014) shows moderate semantic shift extending established trajectories, whereas *The Hierarchy Problem and New Dimensions at a Millimeter* (Arkani–Hamed et al., 1998) shows large divergence, exemplifying paradigm reorientation. All panels are visualized in 2D for exposition; analyses are conducted in the original 768-dimensional space.

For exposition only, we project embeddings to two dimensions to illustrate the semantic triangle and distances and depict the direction-based perspective (Figure 1A). Figure 1C-1H report distributions of both similarity measurements and distance measurements.

---

[1] For the clarity of display, for articles with more than 100 citations, we randomly selected the positions of 100 citing documents for display.

All of these measurements exhibit bell-shaped patterns, but statistical tests reject strict normality, so we do not assume any specific distributional form.

To qualitatively support the interpretation of how these semantic measurements manifest in practice, we visualize two representative publications by plotting their references and citing publications in a 2D distance geometry[2] (Figure 1I-1J). While acknowledging the inherent hindsight bias in evaluating historical milestones, these well-known cases are provided strictly as interpretative support rather than core empirical validation. As a representative case, we examine *Fundamentals of XAFS* (Newville, 2014), a paper that has become an important reference in X-ray absorption spectroscopy. Figure 1I shows that this publication lies between the reference and citation centroids. The citation set in a cluster clearly shifted from the references, suggesting that the paper not only summarized existing spectroscopy but also guided its use in distinct semantic clusters, including astronomy and astrophysics. Figure 1J illustrates *The Hierarchy Problem and New Dimensions at a Millimeter* (Arkani–Hamed et al., 1998). Here this publication is located far from the reference centroid, while later citing publications only partly follow it. This reflects how the publication pushed into theoretical frontiers with novel ideas on extra dimensions, contributions that strongly shaped debates in particle physics and cosmology.

## 3.3 Classification

To characterize paradigm dynamic patterns, we classify publications based on the semantic measurements. We use $s_{RC}$ as the primary criterion for ranking and classifying focal publications according to their paradigm shift potential. Figure 1B provides a schematic illustration of this logic. Accordingly, we classify publications by the relative position of $s_{RC}$ with respect to $s_{RP}$ and $s_{PC}$:

- **Exploratory type (0.8%):** $s_{RC} < min(s_{RP}, s_{PC})$. In this case, the semantic relatedness between references and citing publications is weaker than their relation to the focal publication, suggesting that the focal publication mediates a shift away from the established paradigm, guiding subsequent research into a new semantic cluster, a rupture expected to yield high disruption.
- **Consolidating type (92.0%)**: $s_{RC} > max(s_{RP}, s_{PC})$. Here, the references–citing publications relation is the strongest among the three, indicating that the focal publication reinforces and stabilizes the existing paradigm, channeling subsequent publication along established trajectories, a continuity expected to be associated with low disruption.
- **Balanced type (7.2%):** $min(s_{RP}, s_{PC}) \leq s_{RC} \leq max(s_{RP}, s_{PC})$. These cases indicate that the focal publication neither fully consolidates the cluster of references nor decisively reorients citing publications. In practice, this may reflect incremental extensions of prior work or cases where a publication departs from its references but subsequent citing publications remain only partially aligned, likely associated with moderate disruption.

[2] For visualization, all embeddings are projected into two dimensions using principal component analysis (PCA). All statistical analyses, however, are conducted in the original 768-dimensional semantic space.

The strong skewed distribution toward the consolidating type in our dataset highlights how rare paradigm-reorienting contributions are, which is also consistent with what Wu et al. (2019) observed. Extensive robustness checks (e.g., bootstrap resampling, alternative centroiding, and pure text embeddings) confirm these inequalities capture stable topological boundaries rather than methodological artifacts. Furthermore, due to the extreme long-tail distribution of paradigms, standard clustering fails to isolate these boundaries, necessitating our rule-based approach (see Figure S4–S5).

## 3.4 Novelty and disruption indicators

To assess the validity of our semantic framework, we benchmark it against two established bibliometric indicators that have been widely used as proxies for paradigm dynamic contributions: novelty and disruption. These indicators provide complementary, citation-based perspectives on how research departs from or reinforces existing knowledge structures. These benchmarks enable us to evaluate the extent to which our semantic measurements capture recognized signals of paradigm shift.

Novelty is assessed by examining the rarity of journal pairs in references of each publication. Following Uzzi et al. (2013), we compute a $z$-score for each pair of journals:

$$z = \frac{f_{ij} - \mu_{ij}}{\sigma_{ij}} \tag{7}$$

where $f_{ij}$ is the observed frequencies of journal pair $(i, j)$ in the references, and $\mu_{ij}$ and $\sigma_{ij}$ are the mean and standard deviation of its frequencies under multiple null models generated using 30 rounds of Monte Carlo reshuffling while preserving yearly citation counts for each publication. Following their framework, we define each focal publication's novelty score (N-value) as the 10th percentile of these $z$-scores. This choice ensures that novelty is driven by highly atypical combinations. In this definition, lower (often negative) N-values indicate greater novelty, whereas higher N-values correspond to conventional knowledge integration.

Disruption is measured using the disruption index (D) introduced by Funk and Owen-Smith (2017) and further formalized in subsequent work (Wu et al., 2019). For a focal publication $p$, let $N_i$ denote the number of later publications citing $p$ but not its references, $N_j$ the number of publications citing both $p$ and its references, and $N_k$ the number of later publications published after $p$ that cite only its references. The D value is defined as:

$$d = \frac{N_i - N_j}{N_i + N_j + N_k} \tag{8}$$

The correlation among our proposed measurements and aforementioned bibliometric indicators is shown in Figure S6.

## 3.5 Regression model

To evaluate the explanatory power of semantic measurements for citations and disruption over time, we estimate panel regressions within successive five-year windows. We employ quantile regression models to provide robust estimates of the tendency. While novelty serves as an antecedent disturbance inherent to the publication itself, we treat citation and disruption percentiles as two distinct realized outcomes of the subsequent knowledge diffusion process. Specifically, for each publication $p$ in year $t$, we model two outcomes: (i) its citation percentile within year $t$ and (ii) its disruption percentile within year $t$. The baseline regression specification is:

$$Y_p = \alpha_t + \beta_{1t} s_{RP,p} + \beta_{2t} s_{PC,p} + \beta_{3t} s_{RC,p} + \gamma \boldsymbol{X_p} + \varepsilon_{p,t} \quad (9)$$

where $Y_p$ denotes the dependent variable for $p$; $s_{RP,p}$, $s_{PC,p}$, and $s_{RC,p}$ represent the three semantic measurements; and $\boldsymbol{X_p}$ is a vector of controls including funding grants, reference count, publication year, team size and author status (measured by the maximum prior citation percentile among co-authors), with additional controls for journal impact and funding types yielding consistent results (see Supplementary Note 3 and Tables S2–S4) (Boyack & Börner, 2003; Fortunato, Bergstrom, Börner, Evans, Helbing, Milojević, Petersen, Radicchi, Sinatra, Uzzi, et al., 2018; Park et al., 2023; Sarigöl et al., 2014; Tham, 2023).

# 4 Results and Discussion

## 4.1 Linking semantic categories with disruption

We next examine the disruption characteristic of the three semantic types. Disruption does not directly reflect semantic change, but if our framework captures meaningful paradigm dynamics, exploratory, balanced, and consolidating publications should display systematically different disruption patterns.

Figure 2A presents boxplots of $d$ across the three types. The exploratory type has the widest interquartile range and a greater median value, while the consolidating type is tightly centered around low values. The balanced type falls in between, with moderate variation. Figure 2B shows the distributions of $d$ on a log-transformed scale. The density curves again separate: the exploratory type leans toward positive values clearly, the balanced type is mixed, and the consolidating type concentrates heavily near or below zero. Quantitatively, the percent of publications with $d > 0$ is 42.3% in the exploratory type, 33.1% in the balanced type, but only 14.6% in the consolidating type.

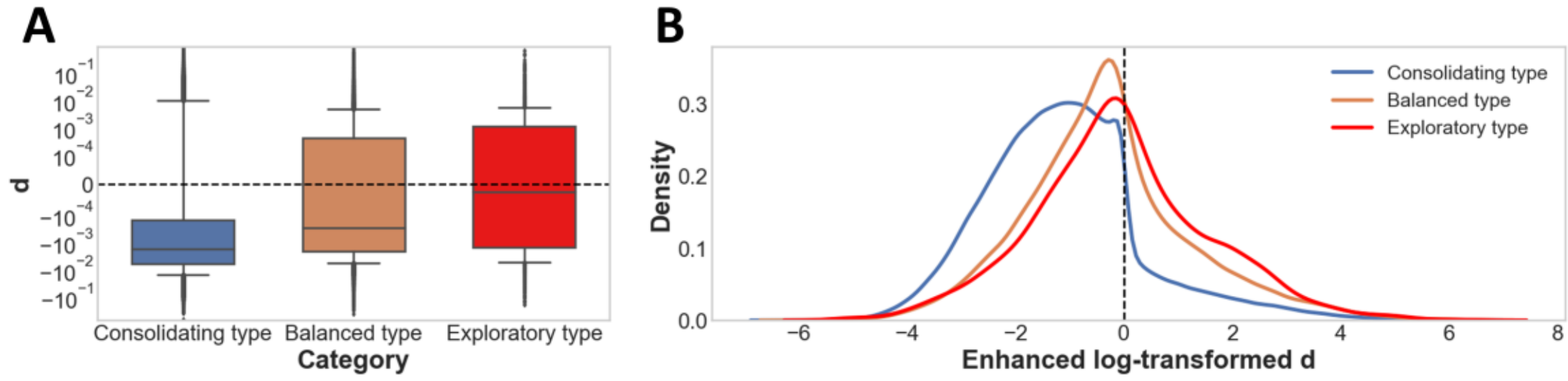

**Figure 2. Distributions of the disruption index across exploratory, balanced, and consolidating publications, visualized as (A) boxplots and (B) density plots (with signed logarithmic transformation [3]).**

These contrasts reveal a consistent ordering, i.e., exploratory>balanced>consolidating, both in median values and in the likelihood of positive disruption. Substantively, this means that publications positioned at the semantic frontier are not only boundary-breaking in semantic space but also more likely to reorient citation flows away from their predecessors. The clear separation across disruption outcomes shows that the semantic classification is not merely formal but reflects substantive differences in how publications consolidate or reorient scholarly trajectories.

This geometry provides a semantic interpretation for disruption, demonstrating that the alignment between references and citing publications ( $s_{RC}$ ) reveals whether a publication reorients knowledge flows. In our framework, a low $s_{RC}$ (or, similarly, a large $d_{RC}$) means that the citation set $\mathcal{C}$ settles in a semantic cluster that is distant from the reference set $\mathcal{R}$, characterizing the publication as more likely to be an exploratory type. This geometry has a direct, structural implication for citation networks. As $\mathcal{C}$ moves away from $\mathcal{R}$, the probability that later publications co-cite the focal publication with its references declines, because the reference set becomes less semantically pertinent to the problems and terminology that compose $\mathcal{C}$. In the disruption decomposition, this shift raises $N_i$ (citing publications to the focal publication that do not cite its references) and lowers $N_j$ (citing publications that cite both the focal publication and its references) per the assumptions of citation analyses in bibliometrics (White & Griffith, 1981). By contrast, when $s_{RC}$ is high and citing publications remain close to the cluster of references, subsequent publications continue to integrate the focal publication alongside its predecessors, sustaining high $N_j$ and producing low or negative disruption. The focal publication may depart from its references ($s_{RP}$ lower) yet still lead followers back toward the same reference neighborhood ($s_{RC}$ higher), preserving co-citation with earlier publication and limiting disruption. Only when the continuity between $\mathcal{R}$ and $\mathcal{C}$ breaks—that is, when $s_{RC}$ collapses—does the co-citation structure reorganize, and the disruption index rise. In this case, $s_{RC}$ captures the semantic dimension underlying disruption: it is the reconfiguration of knowledge flows that makes disruption more than a citation count ratio, grounding it instead in the reorientation of meaning across paradigms.

---

[3] We apply a signed logarithmic transformation, $\tilde{d} = sign(d) \cdot log(1 + 1000|d|)$. This approach preserves the direction of effects while stabilizing variance and highlighting relative differences.

## 4.2 Citation impact across semantic measurements

Citations have long been the primary measurement for scholarly recognition and thus serve as a natural benchmark for evaluating how shifts in paradigm roles map onto scientific impact (Waltman, 2016). We next examine how citation impact varies with reference–citation alignment by evaluating outcomes using both continuous measurements ($s_{RC}$ and $d_{RC}$) and the three semantic types.

Figure 3A shows that citations differ markedly across the three types. The consolidating publications (higher $s_{RC}$) have the highest median and longer upper tails; the exploratory type (lower $s_{RC}$) sits lower and is more dispersed; the balanced type lies between. This initial comparison indicates that semantic continuity with the knowledge base is associated with higher citation impact. Figure 3B shows that the share of consolidating publications among the Top-N% most-cited steadily exceeds a random baseline (i.e., the $N$ value itself) from Top-0.1% to Top-10%. At every threshold, the percentage of exploratory publications is consistently lower, while balanced cases fall in between.

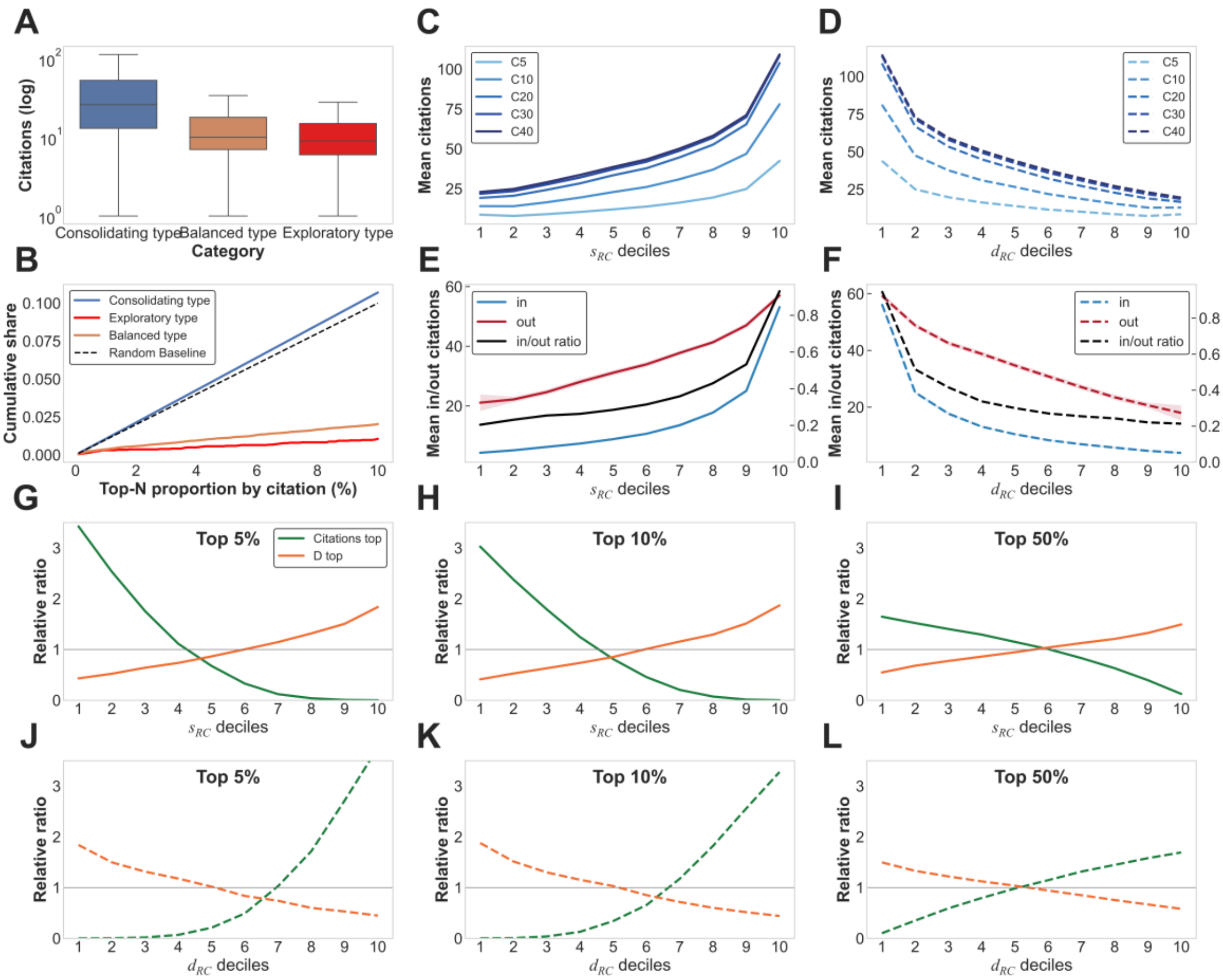

**Figure 3. Citation impact across the semantic measurements**. (A) Distributions of citation counts (log scale) across the three types. (B) Percent of publications in each type represented among the Top-N% most cited, computed at every 0.1% increment. (C–D) Mean citations across $s_{RC}$ and $d_{RC}$ for observation windows ranging from 5 to 40 years. (E–F) Within-field and cross-field citations across deciles of $s_{RC}$ and $d_{RC}$. (G-L) Probability of achieving high citation impact (Top-5%/10%/50%) and high disruption conditional on $S_{RC}$ and $d_{RC}$ deciles (controlling for publication year and field).

Given the inverse relationship between similarity and distance, mean citations rise monotonically with $s_{RC}$ and decline as $d_{RC}$ grows (Figures 3C–3D). This pattern is characterized by a pronounced surge at high similarity and a steep drop at large distance, and remains stable across observation windows ranging from 5 to 40 years.

Figure 3E–3F distinguish between within-field citations [4] and cross-field citations, coming from publications not assigned to astronomy. Both types of citations increase monotonically with $s_{RC}$ and decrease with $d_{RC}$, indicating that greater semantic alignment with the cluster of references enhances visibility both within and beyond the field. Cross-field citations dominate in absolute terms across all deciles, reflecting the porous boundaries of astronomy and its frequent intersections with adjacent areas such as physics, engineering, and data science. Yet the relative growth rate of within-field citations is steeper, such that the in/out ratio rises with $s_{RC}$.

Joint outcomes with disruption are shown in Figure 3G-3L. When we decile publications by $s_{RC}$, controlling for publication year and field, the probability of entering the Top-5%/10%/50% by citations rise sharply with semantic alignment, while the probability of landing in the top disruption bins declines; analyzing $d_{RC}$ yields the reciprocal pattern. Taken together with Section 4.1, this reveals two complementary pathways: consolidation, where close references-citations alignment earns broad, steady recognition; and reorientation, where lower $s_{RC}$ (or higher $d_{RC}$) signals semantic movement away from the cluster of references and is more likely to be disruptive.

Citation counts reflect more than scholarly credit; they are related to the diffusion of knowledge. From this perspective, the geometry of R–P–C helps explain why citation trajectories differ systematically across paradigmatic roles. When $\mathcal{C}$ remains close to $\mathcal{R}$ (higher $s_{RC}$), the citation publications continue to operate within the established paradigm, where shared terminology, methods, and problem framings are already familiar to researchers. This lowers the costs of comprehension and adoption, explaining our finding of rapid recognition, larger familiar audiences, and ultimately higher cumulative citations. By contrast, when $\mathcal{C}$ diverges far from $\mathcal{R}$ (lower $s_{RC}$), subsequent research shifts into less similar semantic cluster, with the focal publication acting as the bridge that facilitates this movement (Leydesdorff, 2007). This introduces what Kuhn described as the challenge of incommensurability, which we operationalize as paradigm conversion costs: new terminology to be learned, new methodological standards to be established, and new problem framings to be legitimated (Crane, 1972; DiMaggio et al., 1983; Kuhn, 1962). In our empirical results, we find that both within- and cross-field citations rise with $s_{RC}$ rising, yet the relative growth is steeper for within-field citations, leading to a rising in/out ratio. Such results imply that these costs slow down diffusion, restrict familiar audiences, and suppress broad recognition (explaining the lower overall citations and selective attention we observe for low $s_{RC}$ publications) (Ash, 2019; Dai, 2020), even when the contribution is conceptually important. These publications often attract attention selectively—especially from adjacent domains that already share some of the new orientation, a selective diffusion pattern consistent with semantic proximity studies (H. Kim et al., 2022)—but accumulate fewer citations overall. Citations thus serve as external evidence that the R–P–C geometry captures not only semantic reconfiguration but also its consequences for how recognition translates into impact.

[4] Within-field citations here refer to those coming from subsequent publications also assigned to the astronomy concept in OpenAlex.

## 4.3 Novelty and semantic measurements

The boxplots in Figure 4A show that publications with high novelty (N-value < 0) exhibit systematically lower similarities across all three pairs $s_{RP}$, $s_{PC}$ and $s_{RC}$ than publications with low novelty (N-value > 0). Figure 4B mirrors this pattern in distance space: larger $d_{RP}$, $d_{PC}$, and $d_{RC}$ under high novelty. Taken together, these contrasts indicate that atypical reference combinations place the focal publication closer to cluster boundaries and are followed by citing publications that drift away from the original cluster of references, i.e., lower $s_{RC}$ as a subsequent outcome.

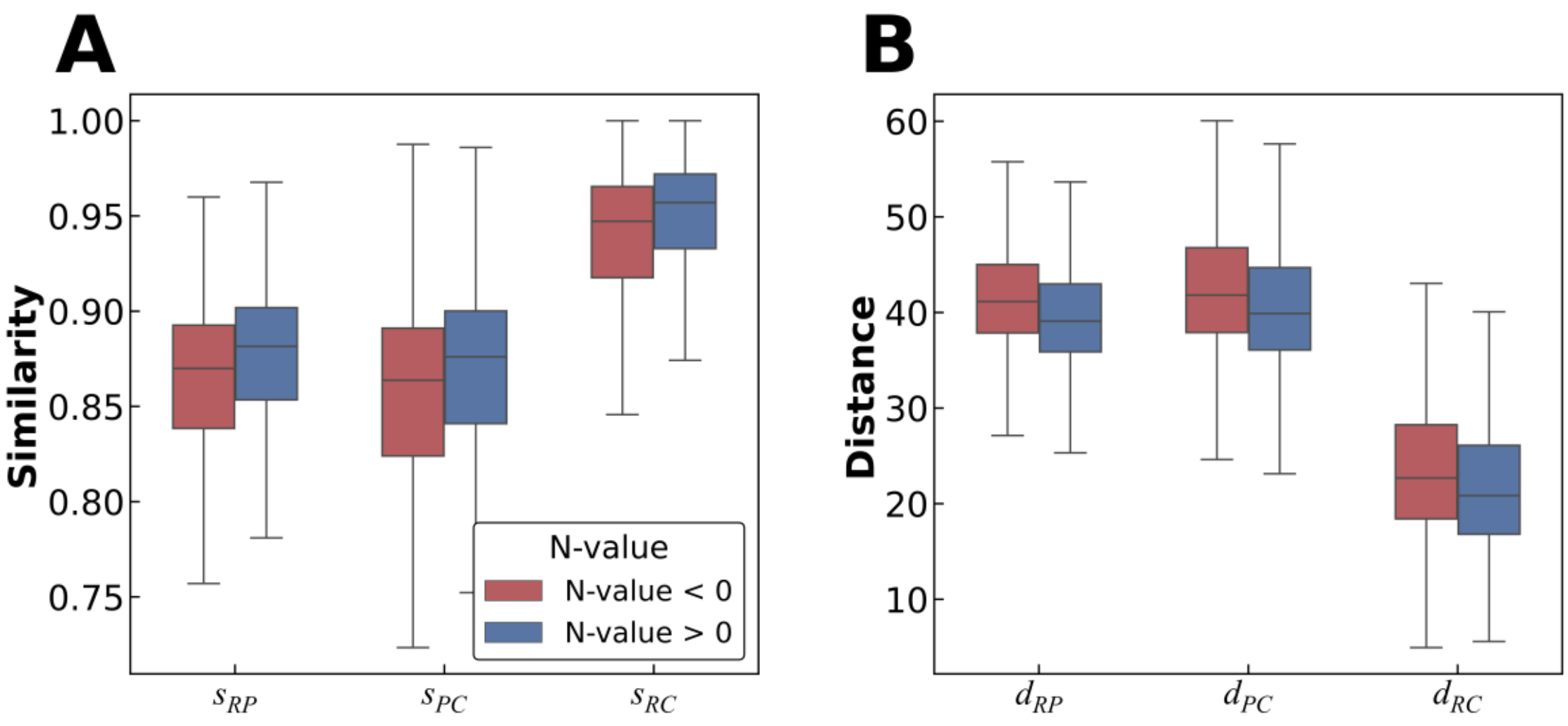

**Figure 4. Novelty and semantic classifications.** Atypical reference combinations predict lower R–P–C similarity. (A-B) Comparison of (A) semantic similarities ($s_{RP}, s_{PC}, s_{RC}$) and (B) Euclidean distances ($d_{RP}$, $d_{PC}$, $d_{RC}$) between publications with high (N-value < 0) and low novelty (N-value > 0).

As shown in Table 1, category shares reinforce this pattern. As the table shows, when novelty is low (N-value > 0), only 0.6% of publications are classified as exploratory and 5.9% as balanced, while the vast majority (93.6%) are consolidating. Under high novelty (N-value < 0), the exploratory share rises from 0.6% to 0.9%—a relative increase of about 50%—and the balanced share grows from 5.9% to 8.0%, a relative increase of about 35.6%. These shifts come at the expense of the consolidating share, which falls modestly from 93.6% to 91.1%. In short, novel reference combinations tilt the distribution away from purely consolidating outcomes and increase the chances of exploratory or balanced types, making reduced $s_{RC}$ the typical signature of novelty.

**Table 1. Percentage of exploratory, balanced, and consolidating publications by novelty score (N-value).**

| Publication type | Consolidating type | Exploratory type | Balanced type |
|---|---|---|---|
| N-value > 0 (Low novelty) | 93.6% | 0.6% | 5.9% |
| N-value < 0 (High novelty) | 91.1% | 0.9% | 8.0% |
| comparatively Δ | -2.7% ($=\frac{91.1\%-93.6\%}{93.6\%}$) | +50.0% | +35.6% |

Our framework thus elucidates the mechanisms underlying novelty in scientific research. Novel combinations of references tend to link terminology and problem framings from distant clusters. Supplementary analyses of within-set semantic dispersion empirically confirm this: highly novel publications systematically draw upon reference sets with significantly higher internal variance (Supplementary Note 5 and Figure S7). Publications with higher novelty consistently show reduced similarity with their references (lower $s_{RP}$), which aligns with prior semantic novelty studies (Shibayama et al., 2021), and with their subsequent citing publications (lower $s_{PC}$). Most critically, they also depress $s_{RC}$, because later publications inherit the heterogeneity of the knowledge base and extend it further into new semantic cluster. In this sense, novelty provides the antecedent condition (the initial disturbance), and it is through the R–P–C geometry that we can observe its semantic consequences: lower $s_{RP}$, $s_{PC}$, and especially $s_{RC}$ trace how unconventional reference combinations trigger semantic realignment and open the pathway to paradigm reorientation (Constantino et al., 2025; S. Yang & Youn, 2024).

## 4.4 Team size and paradigmatic roles

We next examine the social dimension of paradigm dynamics by analyzing team size. Our findings reveal a strong association between collaborative scale and paradigmatic role. Specifically, larger collaborations systematically align with paradigmatic consolidation, whereas smaller teams preserve a higher likelihood of exploratory departures.

In **Figure 5**A–5B, we see that mean similarities $s_{RP}$, $s_{PC}$, and, especially, $s_{RC}$ rise steadily with team size, while the corresponding distances ($d_{RP}$, $d_{PC}$, and $d_{RC}$) decline in parallel, leading to a contraction of the semantic triangle's area. Together, these shifts indicate that larger collaborations tend to situate their work more tightly within the semantic space defined by their references and subsequent citing publications, embedding their contributions more firmly in established paradigms.

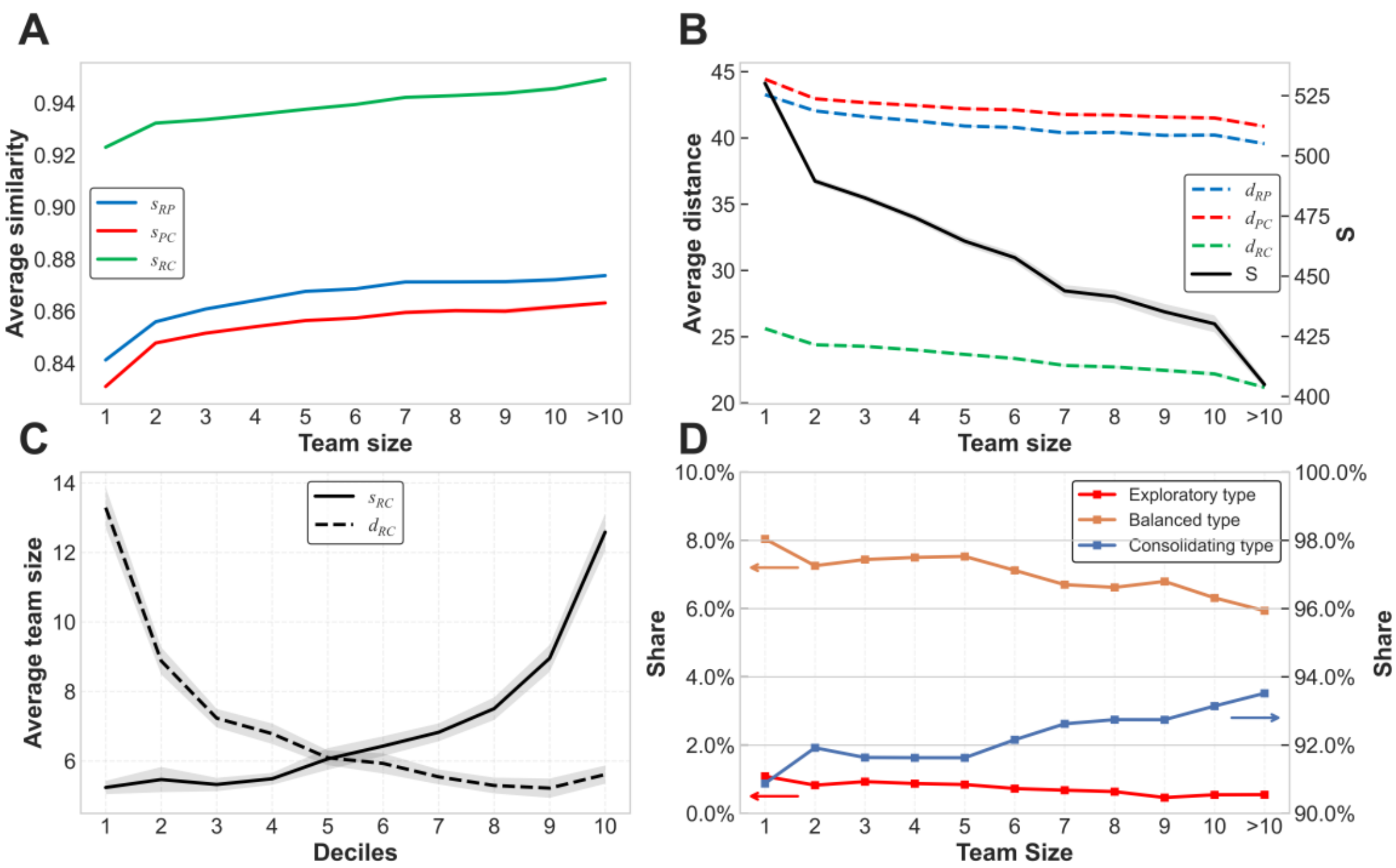

**Figure 5. Team size and paradigmatic roles**. (A–B) Mean semantic similarities ($s$) and Euclidean distances ($d$) plotted against team size. (C) Average team size across deciles of $s_{RC}$ and $d_{RC}$. (D) Distribution of consolidating, balanced, and exploratory types across different team sizes.

Turning to **Figure 5**C, the decile analysis of $s_{RC}$ and $d_{RC}$ further clarifies this association. Publications with higher $s_{RC}$ are produced, on average, by larger teams, while those with higher $d_{RC}$ are associated with smaller teams. The systematic concentration of large collaborations in the high-similarity/low-distance range suggests that semantic conservatism is not incidental but closely tied to team scale (Kamal & Baten, 2025; Lin et al., 2021; Wu et al., 2019).

Finally, **Figure 5**D shows how the distribution of classifications shifts with team size. The share of exploratory types steadily diminishes as the number of authors increases, whereas consolidating cases become more frequent. Balanced cases, by comparison, exhibit less sensitivity to team size fluctuations. These distributions reinforce the broader pattern: larger teams are systematically more likely to contribute to the consolidation of paradigms, while smaller teams preserve a higher potential for pioneering new semantic directions. This reflects the structural logic of collaboration: as the number of co-authors grows, the costs of communication and coordination increase (Sonnenwald, 2007), which incentivizes reliance on familiar terminologies, standardized methods, and canonical problems. Large collaborations thus achieve efficiency and reliability by situating their contributions firmly within established knowledge base. Smaller teams invert this profile: greater latitude to experiment lowers $s_{RP}$, their citing publications are less tightly aligned ($s_{PC}$ weaker), as emerging concepts often lack a fixed interpretation, inviting diverse understandings; and references and citations drift apart ($s_{RC}$ lower), enlarging the triangle and producing paradigm reorientation. Here, the reduced need for consensus allows smaller groups to tolerate higher risks and explore peripheral problem framings, thereby generating unconventional combinations. As a result, they can more readily detach from the cluster of references and guide citations into novel semantic cluster, even at the cost of lower immediate recognition—a pattern consistent with the disruptive role of small teams (Wu et al., 2019).

## 4.5 Temporal dynamics of paradigm formation and differentiation

Figure 6 summarizes how semantic measurements evolve across time and how these shifts condition citation impact and disruption. Figure 6A shows that the annual number of publications has grown with the expansion of astronomy since 1960, yet exploratory cases remain rare. In Figure 6B-6C, the semantic measurements exhibit distinct temporal patterns: $s_{RP}$ rises steadily from the 1950s onward, while $s_{PC}$ and $s_{RC}$ climb up to about 2005 (partly attributable to the time window) and decline thereafter. The distance measurements mirror these trends: while $d_{RP}$ continues a gradual downward drift, $d_{PC}$ and $d_{RC}$ fall until 2005 and then rise. Consistently, the $S$ contracts through 2005 and rebounds afterward.

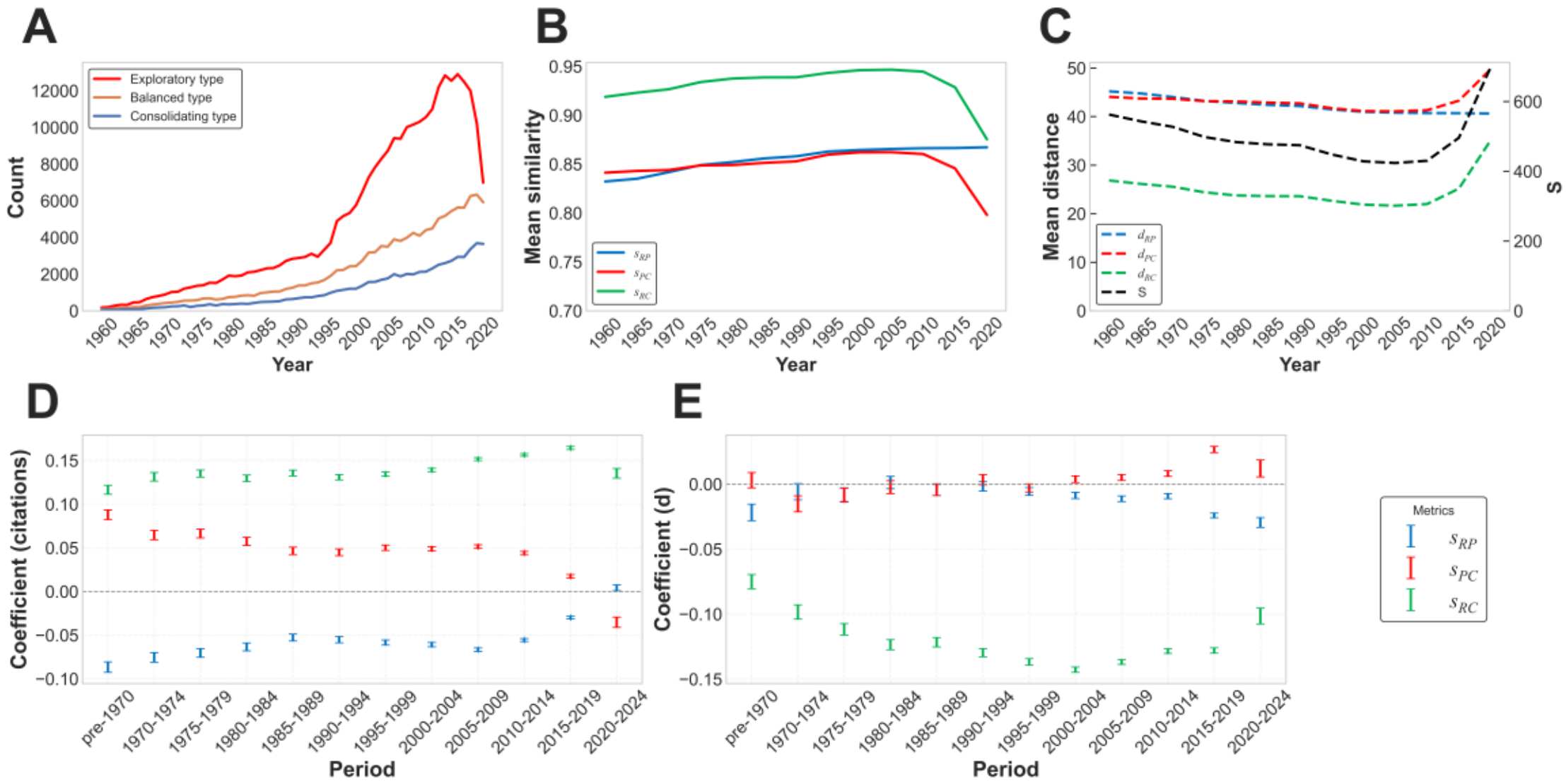

**Figure 6. Temporal dynamics of paradigm formation and differentiation: five-year trends and period-specific regressions.** (A) The annual number of exploratory, balanced, and consolidating publications since 1960. (B–C) Temporal evolution of mean semantic similarities ($s_{RP}, s_{PC}, s_{RC}$) and distances ($d_{RP}, d_{PC}, d_{RC}$) averaged over five-year intervals. (D–E) Period-specific regression coefficients for citation and disruption percentiles with respect to $s_{RP}, s_{PC}$, and $s_{RC}$.

The regression results, as shown in Figure 6D-6E, reveal two patterns. For citations, coefficients on $s_{PC}$ and $s_{RC}$ are consistently positive—strongest around the mid-2000s—whereas $s_{RP}$ is small and often negative. In other words, recognition depends less on the publication's own proximity to its references ($s_{RP}$) than on the continuity between its references and citing publications ($s_{RC}$). For disruption, coefficients on $s_{RC}$ are uniformly negative with the largest magnitude among the three variables, while $s_{PC}$ trends slightly positive and $s_{RP}$ hovers near zero. This indicates that, at every stage, high $s_{RC}$ corresponds to consolidation, whereas low $s_{PC}$ signals the high possibility of semantic reorientation. Crucially, baseline model comparisons (Table S5) demonstrate that integrating these semantic measurements significantly improves the pseudo $R^2$ over models relying solely on traditional institutional controls. This confirms that the semantic geometry provides meaningful new information about paradigm dynamics that cannot be captured by citation topology alone.

The temporal patterns of our semantic measurements reveal how the typical geometric characteristics of publications have evolved over time. The steady rise of $s_{RP}$ indicates that, over time, new publications have become more firmly anchored in their knowledge base: the terminology, methods, and problem framings of astronomy were progressively codified, reflecting the consolidation of a shared disciplinary language. By contrast, the trend of $s_{PC}$ and $s_{RC}$—increasing until around 2005 and declining thereafter—captures a more complex process. In the formative decades, citations extended along similar trajectories as references, signaling a phase of paradigm formation in which the field coalesced around unified clusters. The subsequent downturn reflects differentiation: as astronomy expanded, citations increasingly diffused into distinct semantic neighborhoods, a branching dynamic previously identified by main path (Hummon & Dereian, 1989) and topic evolution models (Blei, 2012), producing divergence between the knowledge base and knowledge diffusion. While astronomy comprises semi-autonomous subareas, this aggregate trend towards differentiation reflects the macroscopic accumulation of asynchronous, localized semantic reorientations rather than a single field-wide shift. To ensure this post-2005 trend is not an artifact of truncation bias (i.e., recent papers having shorter citation windows), we rigorously validated these dynamics using strict fixed-window constraints (e.g., fixed 2-year and 5-year citation windows, right-censoring recent years accordingly). The temporal shift towards differentiation remains highly robust (see Supplementary Note 2 and Figure S8–S9). This trend captures the dual character of paradigm development: integration within a stable canon coexisting with the branching of new specialties. The R–P–C geometry thus provides a semantic trace of paradigm dynamics, capturing the fusion of ideas with references and the subsequent diffusion of knowledge through citations.

## 5 Conclusions

In this paper, we define the semantic geometry of each publication through three pairwise relations: (i) R-P (references–focal publication), which gauges how far the focal publication departs from its inherited terminology and problem set; (ii) P-C (focal publication–citing publications), which records the direction in which subsequent research extends relative to the focal publication; and (iii) R-C (references–citing publications), which captures whether the knowledge diffusion remains anchored to the knowledge base or moves into a different semantic cluster. Measured in Euclidean distance (separation) and cosine similarity (directional alignment), the R–P–C geometry of a publication yields an interpretable scale of contribution to paradigm dynamics: large R–C separation (lower $s_{RC}$) indicates that the focal publication helps pull attention into a new semantic cluster, whereas smaller R–C separation (higher $s_{RC}$) indicates consolidation within the inherited cluster. Therefore, our three semantic types characterize a publication's role within its local semantic cluster, driving paradigm dynamics at the subfield level rather than encompassing the entire discipline.

Our results show that the continuity between $\mathcal{R}$ and $\mathcal{C}$ ($s_{RC}$) offers a direct lens on this dynamic: high $s_{RC}$ corresponds to consolidation, where knowledge flows remain anchored in existing clusters, while low $s_{RC}$ signals reorientation, where attention diffuses into new semantic clusters. This semantic interpretation aligns with and explains multiple established indicators. Disruption rises precisely when $s_{RC}$ collapses, because citations follow the focal publication but bypass its references, thereby breaking the structural link to the knowledge base. Citations themselves reflect the impact of adoption costs: publications close to their clusters of references (higher $s_{RC}$) diffuse rapidly and accumulate broad credit, while those further away face slower, more selective adoption despite their potential to reorient fields. Novelty, constructed *ex ante* from atypical journal pairings (Liu et al., 2024), consistently predicts lower $s_{RC}$, showing how unconventional combinations at the point of synthesis foreshadow semantic reorientation downstream. Finally, the long-run history of astronomy illustrates the evolutionary trajectory Kuhn anticipated: an early phase of paradigm formation, in which $s_{RP}$, $s_{PC}$, and $s_{RC}$ rise as terminology converge, followed by a phase of differentiation, in which the decline of $s_{PC}$ and $s_{RC}$ signals the branching of specialties and the fragmentation of a once-unified paradigm (Kuhn, 1962; Xu et al., 2018). Collectively, these patterns demonstrate that our semantic framework does not stand apart from Kuhn's account but grounds it empirically, showing how semantic relatedness between references and citing publications of a publication capture the mechanisms through which paradigms consolidate, diverge, and evolve. Importantly, supplementary analyses confirm that these structural dynamics and performance hierarchies hold universally across highly diverse disciplines, including Genetics, Epidemiology, Mechanical Engineering, and History. This confirms cross-disciplinary generalizability (Figure S10).

Our findings extend the utility of citation signals, situating them as a semantic lens for uncovering the structural reconfiguration of meaning. Disruption is informative about structural rewiring, but its meaning becomes clear only when paired with the R–P–C geometry, especially $s_{RC}$ that directly gauges whether the knowledge diffusion departs from the knowledge base. In practice, $s_{RC}$ offers an interpretable complement to network indices: high $s_{RC}$ flags consolidation that communities can absorb quickly and widely; low $s_{RC}$ flags reorientation that diffuses more slowly because it introduces new terminology, methods, and problem framings. Since $s_{RC}$ reflects the evolving consensus of the citing publications of a publication, it functions as a dynamic monitor for emerging paradigms, allowing stakeholders to track semantic reorientation in real time as citations accumulate. In practice, stakeholders can monitor recent publications for persistently low $s_{RC}$ to flag nascent shifts in terminology, methods, and problem framings; such monitoring can inform frontier scouting. Evaluation and funding can also be calibrated to paradigmatic role. Low-$s_{RC}$ (exploratory) publications typically face higher adoption costs—new terms to learn, tools to build, norms to settle—so assessment should emphasize qualitative markers of intellectual influence (e.g., adoption of terminology across areas, cross-concept citations) and allow longer evaluation windows. High-$s_{RC}$ (consolidating) publications, by contrast, underwrite shared infrastructure and are appropriately rewarded for rapid, broad adoption. In short, embedding publications in their R–P–C semantics makes paradigm dynamics an actionable diagnostic for research governance.

Our framework demonstrates how semantic geometry can illuminate paradigm dynamics. The measurements rely on document-level aggregations. By compressing the entire text into a single vector, our approach does not disentangle distinct aspects of meaning—such as specific terminology versus methodological frameworks—that may evolve differently during a paradigm shift. Moreover, averaging embeddings into a single point may inevitably obscure finer intra-document details and conflicting semantic signals within a publication. Additionally, we acknowledge that our observed dynamics are not purely semantic; practical and sociological constraints, such as large teams requiring consensus for coordination, serve as critical alternative explanations for their consolidating tendencies. We acknowledge that domain-specific citation behaviors, such as astronomy publications referencing methodological instruments distinct from their core scientific focus, can introduce semantic outliers. However, our supplementary analyses (Figure S3) confirm that the R-P-C geometry remains remarkably robust even when replacing simple centroids with outlier-resistant aggregations like medoids and PCA-ABTT dimensionality reduction. Furthermore, our strict filtering criteria may limit the framework's ability to identify publications with delayed recognition, such as "Sleeping Beauties". Future research could, therefore, enrich the framework along both methodological and practical lines. Methodologically, future studies could incorporate multi-layer linguistic features and alternative embedding models to improve semantic resolution. Practically, the framework could be extended to applied and predictive contexts, such as early detection of emerging topics, evaluation of interdisciplinary collaborations, and mapping scientific frontiers in real time. Additionally, expanding the analysis to include preprints, grant proposals, and patents would provide a more comprehensive view of the scientific ecosystem. Together, these directions would extend the framework from a diagnostic tool to a predictive instrument for understanding how paradigms emerge, consolidate, and transform.

## Supplementary Information

Supplementary Information is separated in another file.

## Acknowledgments

The authors are grateful to all members of the Knowledge Discovery Lab at Peking University, the review editor, and three anonymous reviewers for their helpful comments. This work is supported by the Major Project of the National Social Science Foundation of China (#24ZDA078).

## Generative AI Statement

Generative AI was used for language polishing and editing during the writing process. The authors reviewed the final text and remain responsible for its accuracy.

**Supplementary Information** for A Semantic Geometry for Uncovering Paradigm Dynamics via Scientific Publications

## Supplementary Note 1: Robustness of the semantic geometry framework

To ensure that our tripartite classification and the observed citation-disruption dynamics are not artifacts of specific methodological choices, we conducted an extensive series of robustness checks. Specifically, we tested the stability of our findings across four alternative specifications:

- Alternative thresholding for reference and citation inclusion (Figure S11).
- Cross-disciplinary validation across distinct academic fields (Figure S10).
- Alternative representation using text embeddings derived from SciBERT, specifically trained on a curated corpus of 90,463 engineering-field publications (Figure S2).
- Alternative semantic aggregation methods and embedding post-processing (Figure S3).

To address concerns that simple averaging (Simple Mean) might be sensitive to semantic outliers or the inherent anisotropy of vector spaces, we evaluated five alternative definitions of the geometric "center":

- **Geometric median:** Minimizes the sum of Euclidean distances to effectively resist the pull of extreme semantic outliers.
- **Medoid:** Identifies the actual publication vector within the set with the minimum total distance to all others, ensuring the center represents a real-world document rather than a synthetic coordinate.
- **Time-weighted mean:** Applies an exponential decay function ($w = e^{-\lambda\Delta t}$) to weight publications by their recency relative to the core publication.
- **Citation-weighted mean:** Weights references by their historical citation counts at the time of the core publication's publication using a log-transformed weight ($1 + \log(1 + C_{hist})$) to prioritize foundational knowledge.
- **PCA-ABTT (All-but-the-top):** Performs PCA dimensionality reduction and removes the dominant first principal component to eliminate anisotropic artifacts common in high-dimensional embeddings.

Across all four supplementary analyses (Figures S2, S3, S10, and S11), the structural divergence of $s_{RC}$, the performance profiles of the three paradigm roles, and their systemic relationships with novelty and team size remain highly consistent. This confirms that the Semantic Geometry framework captures intrinsic features of the citation network rather than boundary or parameter-dependent artifacts.

## Supplementary Note 2: Robustness against truncation bias and temporal effects

To explicitly address potential truncation bias and right-censoring effects in citation accumulation—particularly the concern that recent publications might exhibit lower similarities simply due to shorter exposure times—we introduced rigorous fixed-window specifications. We recalculated the semantic embedding centers and all corresponding metrics under four standardized constraints: a fixed 2-year observation window, a fixed 5-year observation window, a fixed 5-citation volume window, and a fixed 10-citation volume window.

Figure S8 demonstrates that the core systemic relationships of the Semantic Geometry framework—including the regression coefficients of $s_{RC}$, the performance profiles of the three paradigm roles, and their associations with novelty and team size—are highly stable across these temporal and volume constraints. This confirms that our framework is invariant to the variance in citation accumulation periods.

Furthermore, to verify that the historical shift toward "differentiation" (post-2005) is an authentic paradigm evolution rather than a temporal artifact, we plotted the temporal dynamics strictly using the fixed 2-year and 5-year windows (Figure S9). To guarantee that every publication in the cohort had equal time to accrue citations, we explicitly truncated the analyzed publication years based on the data collection limit (restricting the 2-year window cohort to publications published up to 2018, and the 5-year window cohort to 2015). The persistent divergence in semantic similarities and the continuous shifting proportions of paradigm roles in these controlled subsets confirm that the observed trends represent a genuine structural evolution in modern science, independent of truncation bias.

## Supplementary Note 3: Robustness checks with additional controls

To verify that the explanatory power of semantic indicators is not an artifact of institutional factors, we conducted supplementary regressions. We present these analyses here to maximize sample coverage, as traditional Journal Impact Factors (JIF) exclude conference proceedings and funding records are not universally available. Table S2 reports the global regression results after additionally controlling for the JIF and author status. Furthermore, to account for varying resource contexts, Table S3 and Table S4 present a heterogeneity analysis across four funding categories: Government/Professional (N = 28,940), Academic Institution (N = 31,801), Publisher (N = 20,087), and Unfunded (N = 428,511).

Across all model specifications, the core semantic metrics ($s_{RP}$, $s_{RC}$, and $s_{PC}$) maintain consistent significance and direction. Notably, $s_{RC}$ remains a robust predictor for both impact dimensions—maintaining a positive correlation with citation and a negative correlation with disruption—regardless of funding sources or journal prestige, confirming its independent explanatory power for scientific innovation.

Finally, to explicitly demonstrate that our semantic metrics add meaningful new information beyond traditional structures, we compared the explanatory power of our full specifications against baseline models (which include only the non-semantic control variables). To ensure that this incremental value is truly driven by semantic content rather than any citation-signal leakage, we evaluated this baseline comparison across both our default embeddings and purely text-trained SciBERT embeddings. As shown in Table S5, the integration of the semantic measurements ($s_{RP}$, $s_{PC}$, $s_{RC}$) yields a distinct and robust increase in the pseudo $R^2$ for both citation and disruption outcomes under both embedding modalities.

## Supplementary Note 4: Validation of paradigm role classification

To rigorously validate our rule-based classification (Exploratory, Consolidating, Balanced) and address potential biases from data filtering, we conducted comprehensive stress tests against potential methodological artifacts.

First, we evaluated the sensitivity of our classification proportions to different reference and citation inclusion thresholds (Table S1). While the exact percentages shift predictably—stricter citation thresholds (e.g., # Cit ≥ 15) increase the proportion of consolidating publications and decrease exploratory ones, reflecting the higher adoption costs and selective diffusion of exploratory work—the fundamental highly skewed distribution remains remarkably robust across all subsets. Exploratory publications consistently represent a rare fraction (< 3%), while consolidating publications form the vast majority (> 82%), confirming that our taxonomy is not an artifact of dataset truncation.

Furthermore, as shown in Figure S4, our classifications remain highly stable across alternative semantic aggregations (Geometric Medians, PCA-ABTT) and under 100x bootstrap resampling of citation networks. Notably, when tested using purely text-trained SciBERT embeddings, the absolute discrete label retention drops. However, this serves as a powerful negative control: un-finetuned language models suffer from representation degeneration (anisotropy), heavily compressing the geometric space. While absolute boundaries shift, the relative systemic trends remain entirely robust (see Figure S2).

Furthermore, we evaluated the necessity of our inequality rules against unsupervised learning (Figure S5). Because paradigm dynamics follow an extreme long-tail distribution (~92% consolidating vs. ~0.8% exploratory), standard distance-based K-Means clustering merely bisects the massive density core, completely failing to isolate exploratory innovations (Figure S5B). This confirms that our inequalities correctly anchor onto topological semantic ruptures rather than data density. Finally, the normalized triangle Area $S$ proves to be highly discriminative, with Exploratory publications consistently occupying the highest global percentiles (Figure S5A).

## Supplementary Note 5: Within-set dispersion and multimodal semantic structures

A potential limitation of representing reference ($\mathcal{R}$) and citing ($\mathcal{C}$) sets via simple centroids is that it may obscure highly dispersed or multimodal semantic structures. For instance, a highly novel publication might synthesize knowledge from two distant semantic clusters. Averaging these widely separated clusters could artificially place the centroid in a sparse "middle" region, which might distort distance metrics.

To explicitly test whether our semantic classifications capture genuine divergence rather than centroid collapse artifacts, we analyzed the within-set dispersion (Euclidean variance and mean Euclidean distance) of the reference sets. Across our dataset, the average within-set Euclidean variance is substantially large (2050.9 for Reference sets and 1797.9 for Citing sets), confirming an inherent degree of semantic breadth and high dispersion within these document neighborhoods. Notably, we observe profound systematic asymmetries in this dispersion across paradigm roles:

- Exploratory publications exhibit the highest within-set variance in their $\mathcal{R}$ sets (2160.8) but the lowest in their C sets (1172.0). This structural profile empirically confirms their theoretical function: they synthesize disparate, heterogeneous knowledge bases to break away from existing paradigms. However, due to high paradigm conversion costs, their initial diffusion is narrow and highly selective, resulting in a cohesive, low-variance Citing set.
- Consolidating publications, conversely, show the lowest dispersion in $\mathcal{R}$ sets (2043.6) but the highest in $\mathcal{C}$ sets (1844.9). This reflects their tight integration within established micro-paradigms (cohesive $\mathcal{R}$), which lowers comprehension costs and allows them to diffuse broadly across the discipline (highly dispersed $\mathcal{C}$).

Furthermore, as shown in Figure S7, publications exhibiting high novelty ($N < 0$) inherently possess a significantly more dispersed reference base compared to conventional publications, verifying that atypical journal combinations draw from multi-cluster semantic neighborhoods.

Given the overall high dispersion and severe structural asymmetry across roles, it is critical to ensure our classification is not an artifact of centroid distortion. Importantly, our core geometric framework proves exceptionally robust: when replacing mathematical centroids with empirical Medoids or Geometric Medians (which identify actual geometric centers rather than synthetic middle points), the classification proportions, the performance hierarchies across paradigm roles, and the novelty-lower $s_{RC}$ mechanism remain strictly consistent (see Figure S3). This confirms that our semantic geometry captures genuine paradigm dynamics, independent of potential multimodal centroid collapse.

# Supplementary Figures

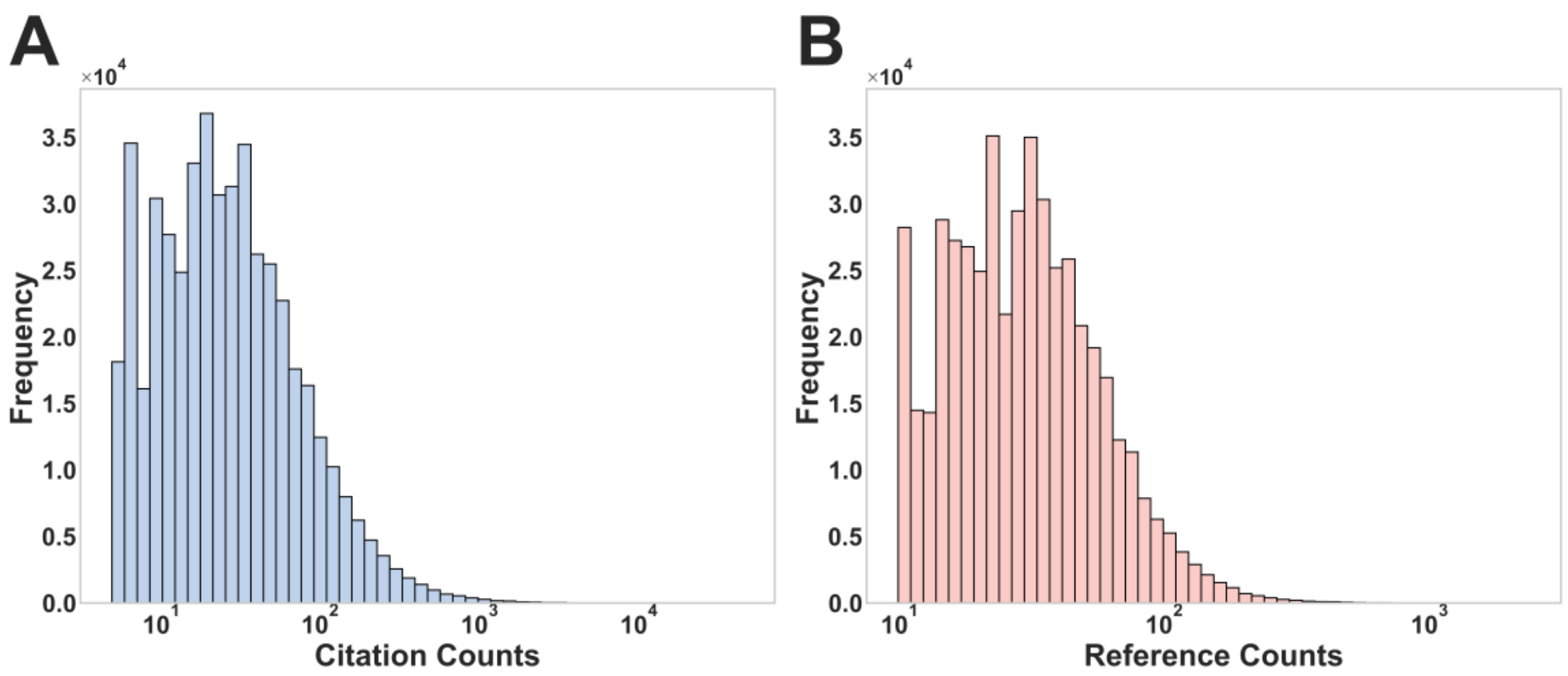

**Figure S1. Distributions of citation counts and reference counts of publications used in this study.**

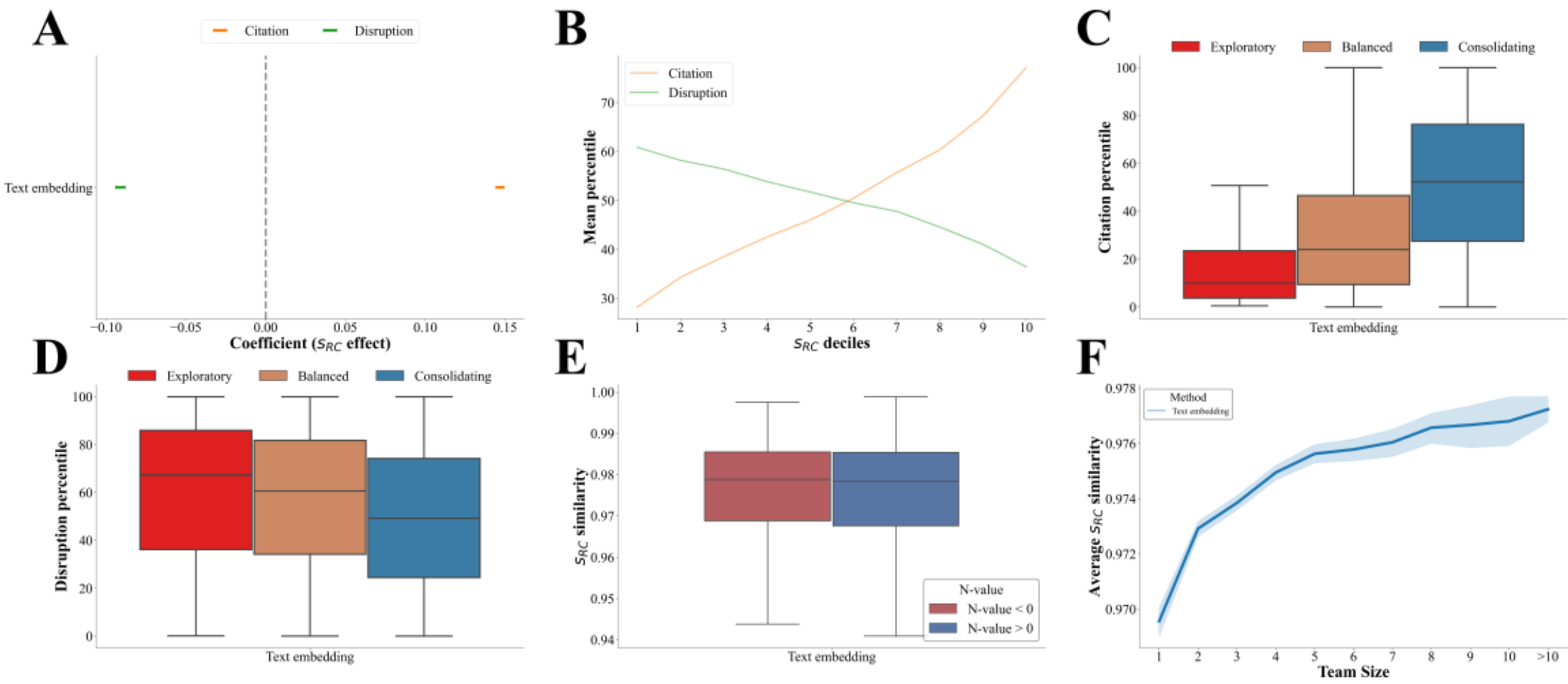

**Figure S2. Robustness check using pure text embeddings.** This figure replicates the core analyses by substituting the original embeddings with pure text-based semantic embeddings derived from SciBERT to construct the semantic geometry. This configuration is designed to isolate textual and linguistic content from citation network structures, testing the framework's sensitivity to the underlying embedding modality. (A) Global regression coefficients of $s_{RC}$ on citation and disruption percentiles. (B) Mean year-normalized percentiles across $s_{RC}$ deciles. (C-D) Boxplots comparing the citation (C) and disruption (D) performance among exploratory, balanced, and consolidating publications. (E) Comparison of $s_{RC}$ distributions between novel (N-value > 0) and non-novel (N-value < 0) publications. (F) Average $s_{RC}$ trends across different team sizes. The highly consistent trends observed within the pure text semantic space demonstrate that our framework is not contingent upon specific network properties, confirming the cross-modal robustness of the scientific evolutionary dynamics it reveals.

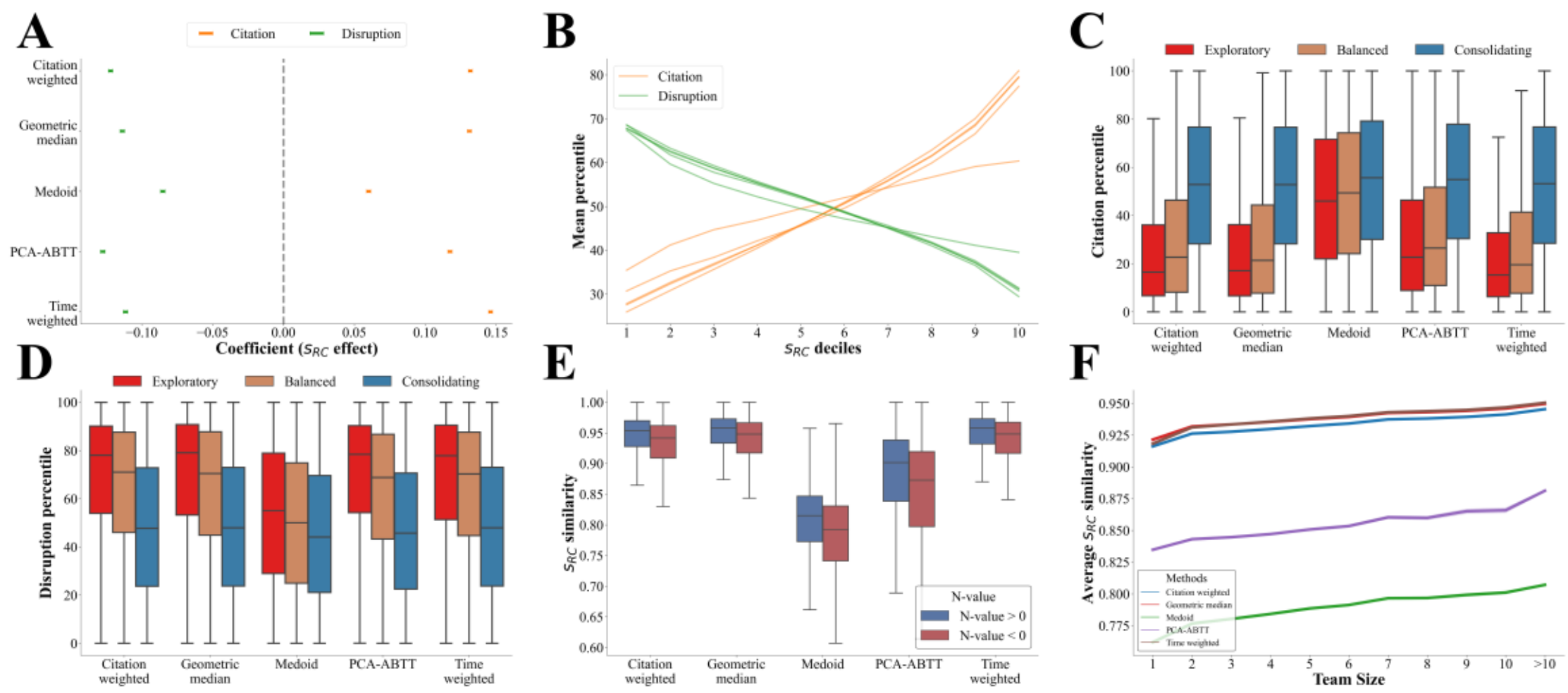

**Figure S3. Robustness checks across alternative semantic aggregation methods and embedding post-processing.** This figure replicates the core analyses using six different specifications: Simple Mean (baseline), Geometric Median, Medoid, Time-weighted Mean, Citation-weighted Mean, and PCA-ABTT. (A) Global regression coefficients of $s_{RC}$ on citation and disruption percentiles. (B) Mean year-normalized percentiles across $s_{RC}$ deciles. (C-D) Boxplots comparing citation (C) and disruption (D) performance among exploratory, balanced, and consolidating publications. (E) Comparison of $s_{RC}$ between novel (N-value > 0) and non-novel (N-value < 0) publications. (F) Average $s_{RC}$ trends across different team sizes. The consistent trends across all geometric definitions confirm that the framework is invariant to specific centroid calculations or high-dimensional embedding artifacts.

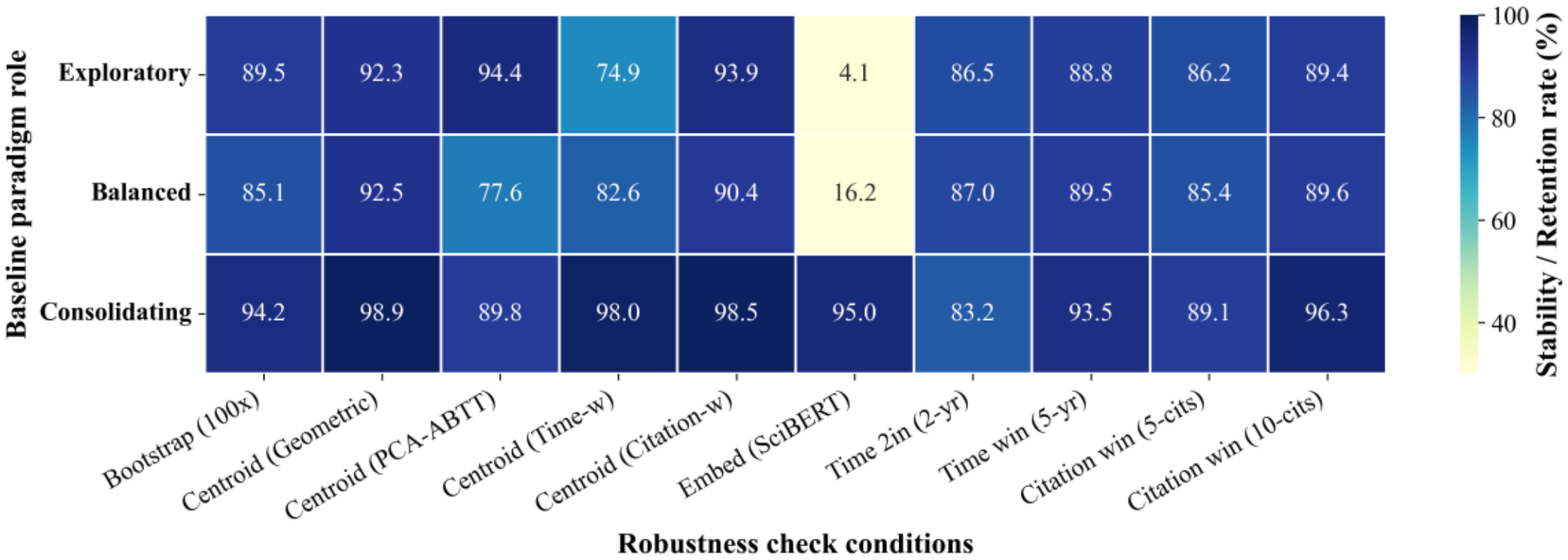

**Figure S4. Classification stability across alternative methodological specifications.** Heatmap showing the discrete label retention rate (%) of publication categories (exploratory, balanced, consolidating) under various stress tests compared to the baseline analysis. Specifications include 100x bootstrap resampling, alternative centroiding algorithms, and cross-modal validation via text-only embeddings.

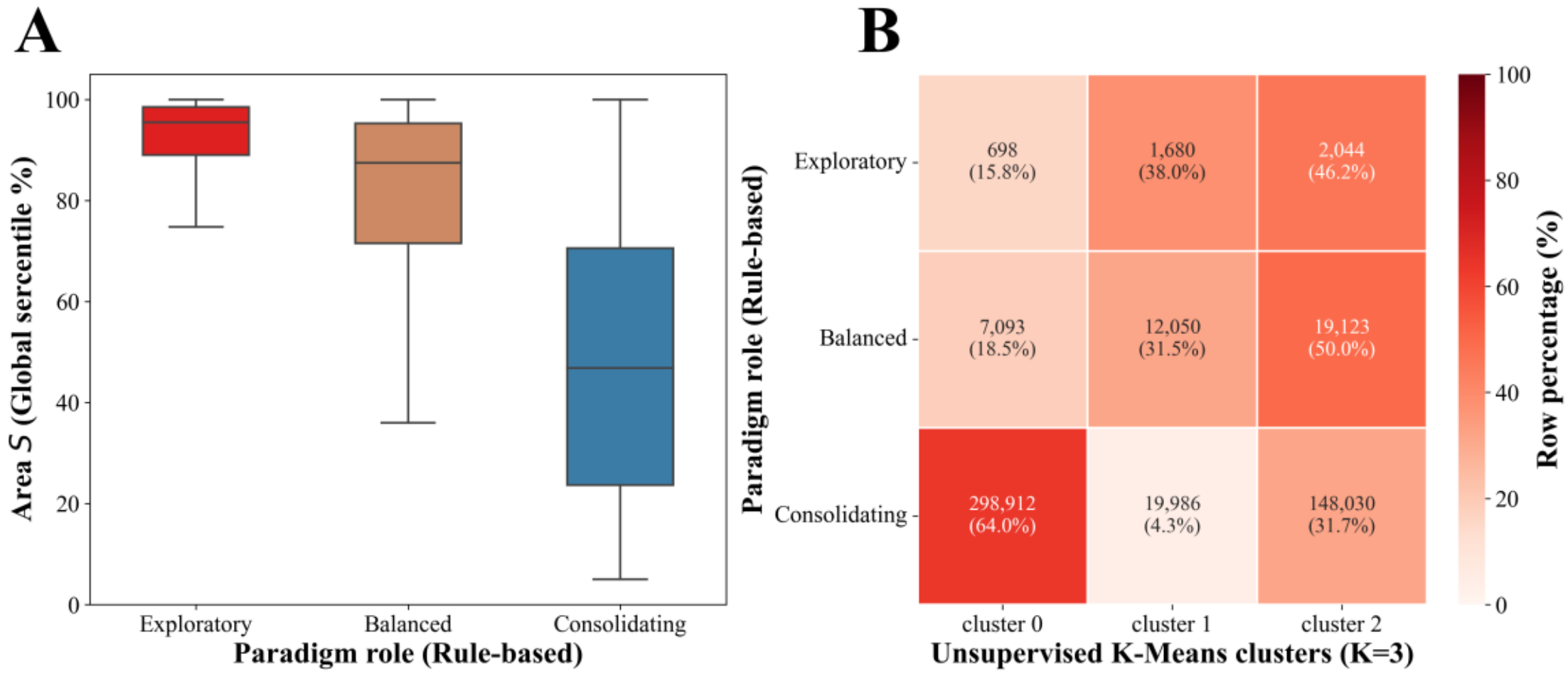

**Figure S5. Geometric validation and cluster analysis of paradigm roles.** (A) Boxplot showing the distribution of normalized semantic triangle Area $S$ (global percentile ranking) for the three classification roles. (B) Confusion matrix demonstrating the composition of unsupervised K-Means clusters (K=3) relative to the rule-based paradigm roles.

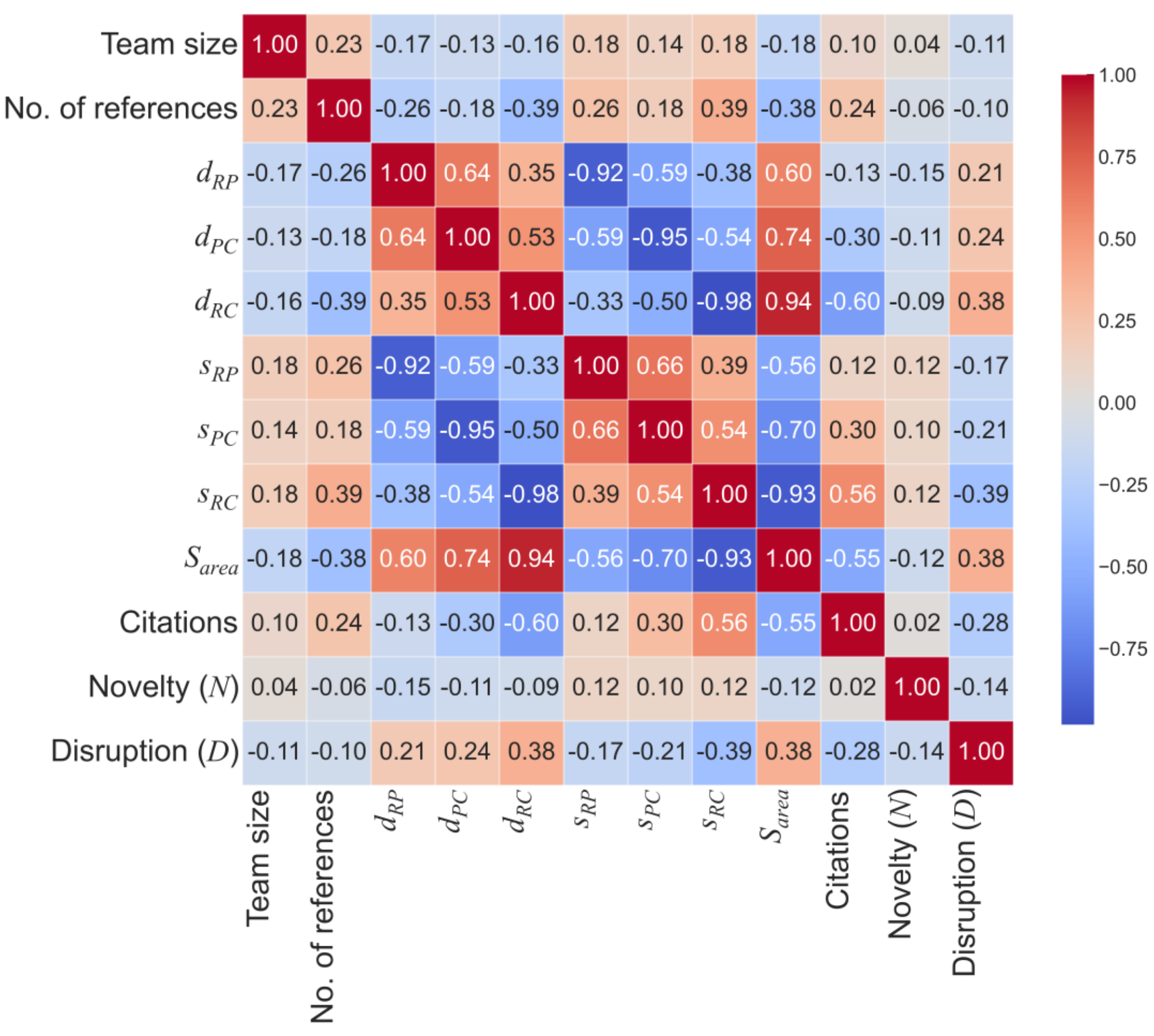

**Figure S6. Spearman correlation matrix of semantic measurements and bibliometric indicators.**

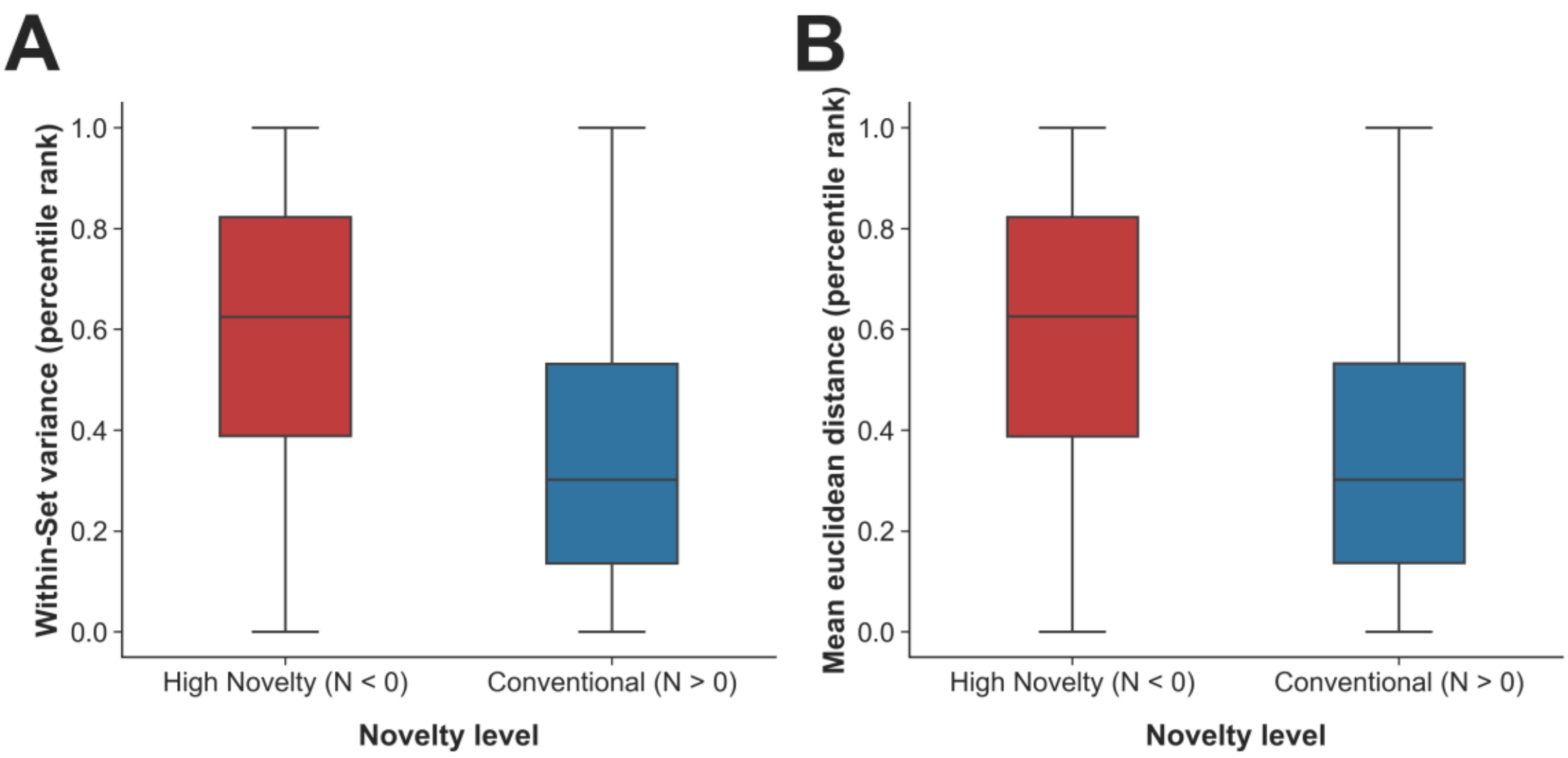

**Figure S7. Within-set semantic dispersion of reference sets across different novelty levels.** To assess the structural dispersion of knowledge bases, we calculated the semantic dispersion within the reference set of each focal publication. Panel (A) displays the within-set Euclidean variance, and Panel (B) displays the mean Euclidean distance (values are presented as percentile ranks). Publications with high novelty ($N < 0$, red) systematically exhibit significantly higher internal dispersion compared to conventional publications ($N > 0$, blue), empirically confirming that atypical reference combinations draw from diverse and highly dispersed semantic neighborhoods.

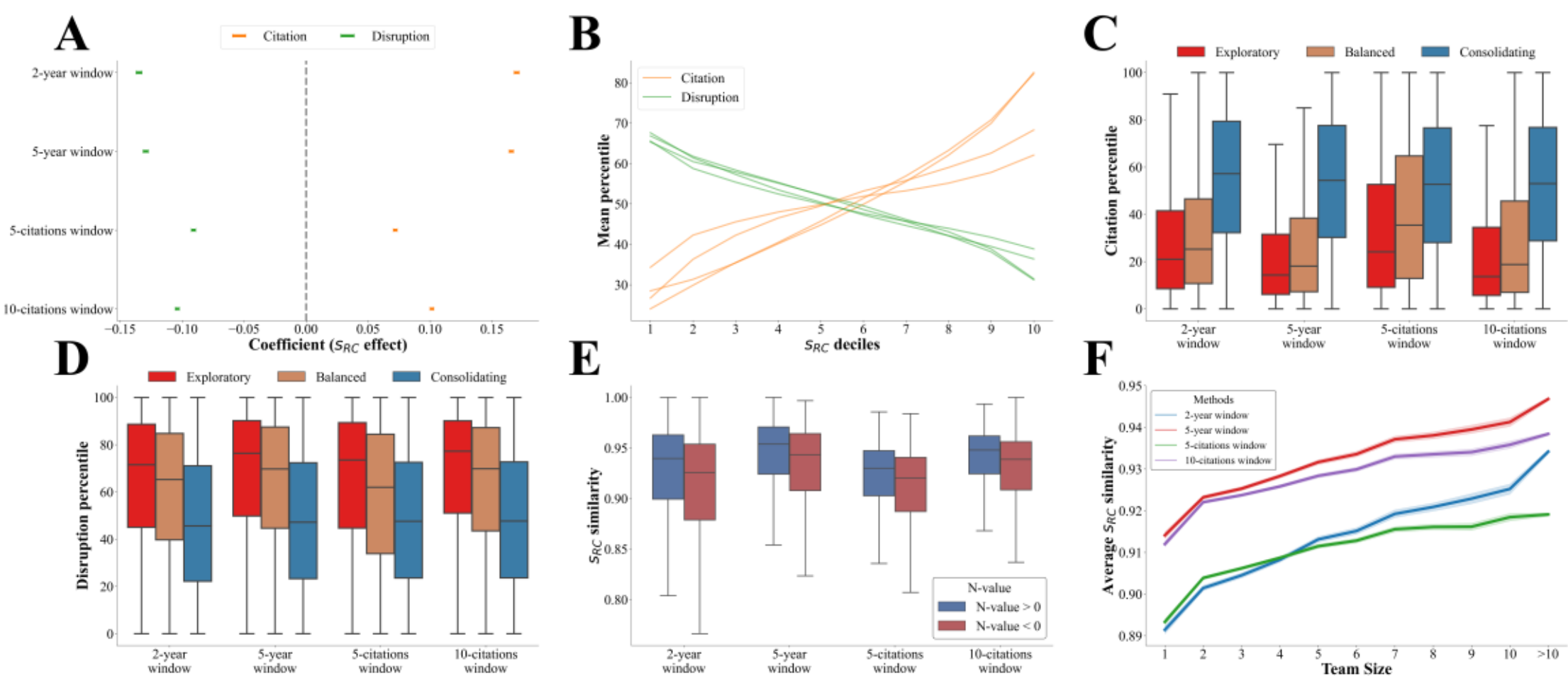

**Figure S8. Robustness checks across fixed temporal and citation-volume windows.** To eliminate potential truncation bias in citation accumulation, this figure replicates the core analyses using standardized observation windows: 2-year time window, 5-year time window, 5-citation window, and 10-citation window. (A) Global regression coefficients of $s_{RC}$ on citation and disruption percentiles. (B) Mean year-normalized percentiles across $s_{RC}$ deciles. (C-D) Boxplots comparing citation (C) and disruption (D) performance among exploratory, balanced, and consolidating publications. (E) Comparison of $s_{RC}$ between novel (N-value > 0) and non-novel (N-value < 0) publications. (F) Average $s_{RC}$ trends across different team sizes. The consistent structural properties across all fixed windows confirm that our findings are not artifacts of varying citation exposure times.

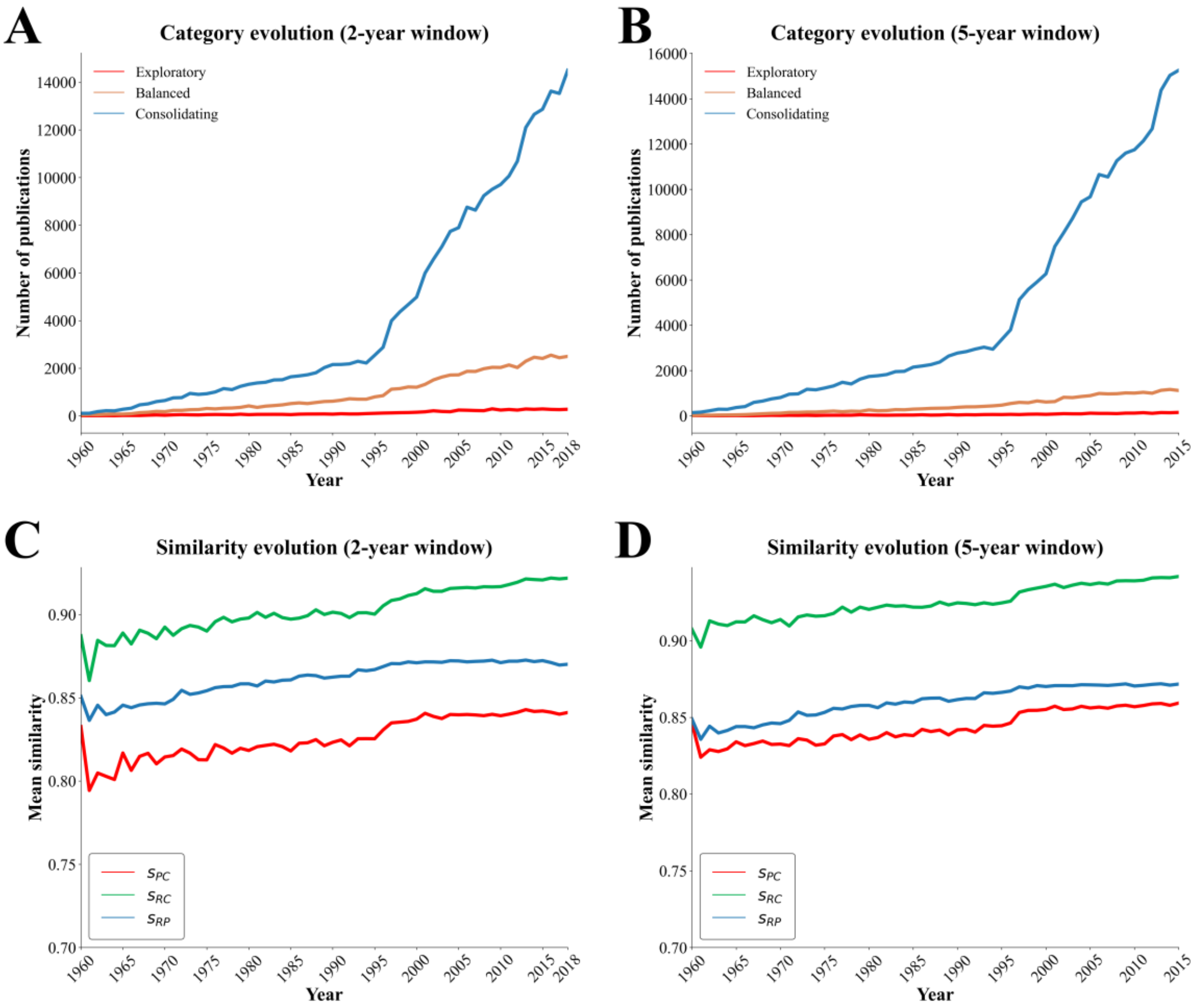

**Figure S9. Temporal evolution of paradigm roles and semantic similarities under strict fixed-window constraints.** To ensure all analyzed publications had strictly equal time to accumulate citations, the timeline is right-censored based on the corresponding observation window (up to 2018 for the 2-year window, and up to 2015 for the 5-year window). (A-B) Category evolution showing the number of publications for exploratory, balanced, and consolidating roles using a 2-year window (A) and a 5-year window (B). (C-D) Evolution of mean semantic similarities ($s_{PC}$, $s_{RC}$, $s_{RP}$) using a 2-year window (C) and a 5-year window (D). The sustained shift toward Consolidating publications and the continuous decline in mean similarities under these strict controls confirm that the historical trend toward differentiation is a real structural phenomenon rather than an artifact of truncation bias in recent years.

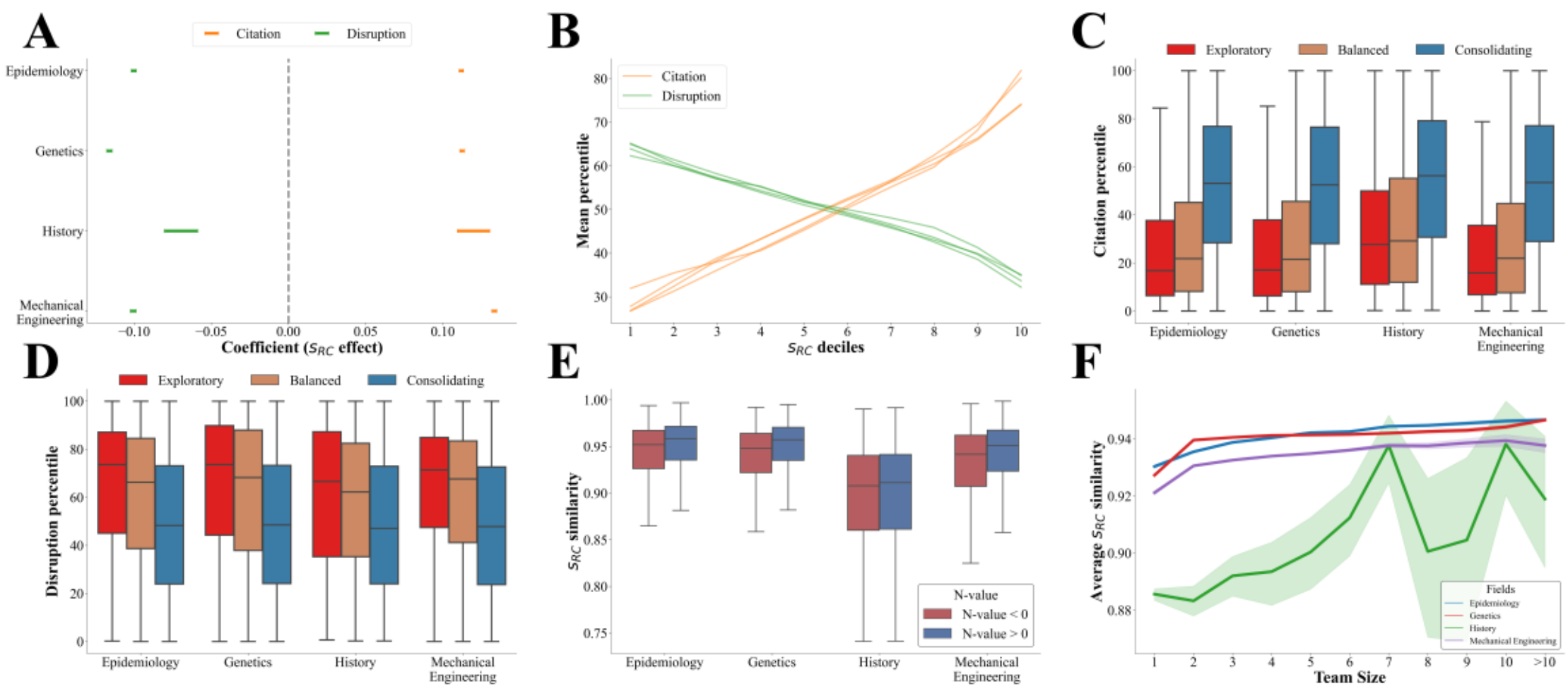

**Figure S10. Robustness checks across distinct academic fields.** This figure replicates the core analyses across four diverse disciplines (Epidemiology, Genetics, History, and Mechanical Engineering) to test the domain sensitivity of the Semantic Geometry framework. (A) Global regression coefficients of $s_{RC}$ on citation and disruption percentiles across the four fields. (B) Mean year-normalized percentiles for citation and disruption across $s_{RC}$ deciles. (C-D) Boxplots comparing citation (C) and disruption (D) performance among exploratory, balanced, and consolidating publications within each field. (E) Comparison of $s_{RC}$ between novel (N-value > 0) and non-novel (N-value < 0) publications across domains. (F) Average $s_{RC}$ trends across different team sizes for each discipline. The persistent structural divergence and consistent performance hierarchies across these highly varied disciplines confirm that the framework captures universal patterns of knowledge progression, independent of domain-specific epistemologies or citation cultures.

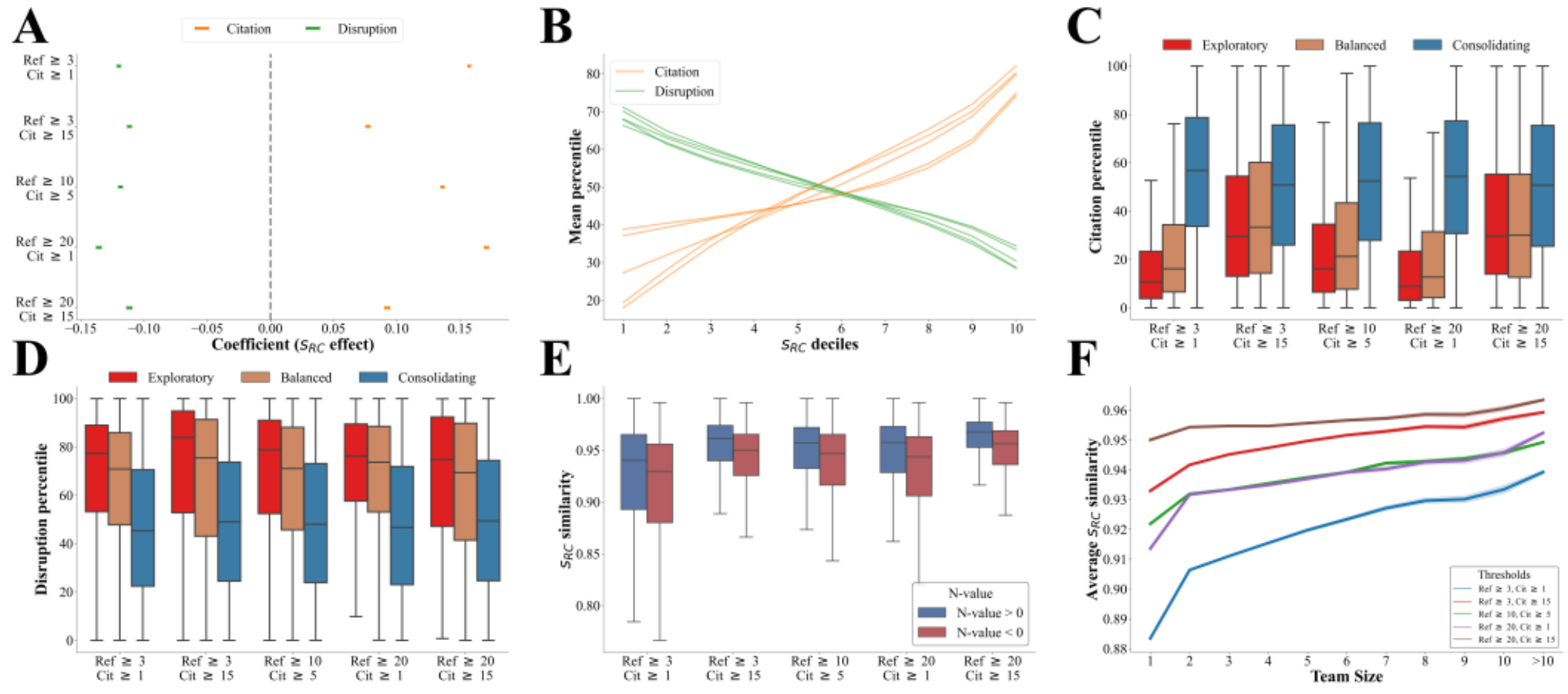

**Figure S11. Robustness checks across varying reference and citation inclusion thresholds.** This figure replicates the core analyses using five increasingly stringent data subsets to test threshold sensitivity. (A) Global regression coefficients of $s_{RC}$ on citation and disruption percentiles. (B) Mean year-normalized percentiles across $s_{RC}$ deciles. (C-D) Boxplots comparing citation (C) and disruption (D) performance among exploratory, balanced, and consolidating publications. (E) Comparison of $s_{RC}$ between novel (N-value > 0) and non-novel (N-value < 0) publications. (F) Average $s_{RC}$ trends across

different team sizes. The consistent trends across all subsets confirm that the framework is invariant to network degree distributions.

## Supplementary Tables

**Table S1. Percentage of exploratory, balanced, and consolidating publications across varying inclusion thresholds.**

| | #Ref≥3, #Cit≥1 | #Ref≥3, #Cit≥15 | #Ref≥10, #Cit≥5 | #Ref≥20, #Cit≥1 | #Ref≥20, #Cit≥15 |
|---|---|---|---|---|---|
| Exploratory | 2.7% | 0.4% | 0.8% | 1.0% | 0.2% |
| Balanced | 14.4% | 4.3% | 7.3% | 9.9% | 2.9% |
| Consolidating | 82.8% | 95.3% | 91.9% | 89.0% | 96.9% |

**Table S2. Global regression results controlling for journal impact and author prestige. Dependent variables are year-normalized citation and disruption percentiles. Standard errors are in parentheses. *** $p < 0.01$, ** $p < 0.05$, * $p < 0.1$.**

| | Citation percentile | Disruption percentile |
|---|---|---|
| $s_{RP}$ | -0.0375*** | -0.0228*** |
| | -0.0011 | -0.0013 |
| $s_{RC}$ | 0.1111*** | -0.1065*** |
| | -0.0011 | -0.0014 |
| $s_{PC}$ | 0.0281*** | 0.0186*** |
| | -0.0013 | -0.0016 |
| # of references | 0.0017*** | 0.0009*** |
| | 0 | 0 |
| Team size | -0.0006*** | -0.0012*** |
| | -0.0002 | -0.0002 |
| # of funders | 0.0239*** | -0.0136*** |
| | -0.001 | -0.0012 |
| $log(jif)$ | 0.1379*** | 0.0033 |
| | -0.0017 | -0.0021 |
| Year author most citation percentile | 0.0778*** | -0.1628*** |
| | -0.0032 | -0.0039 |
| const | 0.1706*** | 0.5831*** |
| | -0.0029 | -0.0035 |
| Observations | 199274 | 199274 |
| Pseudo R-squared | 0.1395 | 0.0652 |

**Table S3. Heterogeneity analysis of citation percentile across funding types. The sample is divided**

**by funding roles. Standard errors are in parentheses.** *** $p < 0.01$, ** $p < 0.05$, * $p < 0.1$.

| | Gov/Prof | Institution | Publisher | Unfunded |
|---|---|---|---|---|
| $s_{RP}$ | 0.0076*** | 0.0067** | 0.0135*** | -0.0388*** |
| | -0.0029 | -0.0027 | -0.0035 | -0.0008 |
| $s_{RC}$ | 0.1293*** | 0.1366*** | 0.1342*** | 0.1328*** |
| | -0.0036 | -0.0034 | -0.0046 | -0.0008 |
| $s_{PC}$ | -0.0077* | -0.0122*** | -0.0139*** | 0.0302*** |
| | -0.0042 | -0.004 | -0.0053 | -0.0009 |
| # of references | 0.0012*** | 0.0013*** | 0.0012*** | 0.0017*** |
| | -0.0001 | 0 | -0.0001 | 0 |
| Team size | 0 | 0 | 0 | 0.0002*** |
| | 0 | 0 | 0 | 0 |
| # of funders | 0.0040*** | 0.0052*** | 0.0037*** | 0 |
| | -0.0007 | -0.0007 | -0.0009 | 0 |
| Year author most citation percentile | 0.1591*** | 0.1584*** | 0.1634*** | 0.1613*** |
| | -0.0085 | -0.0082 | -0.0108 | -0.0021 |
| const | 0.4252*** | 0.3962*** | 0.4331*** | 0.3406*** |
| | -0.0053 | -0.0053 | -0.006 | -0.0013 |
| Observations | 28940 | 31801 | 20087 | 428511 |
| Pseudo R-squared | 0.1351 | 0.1467 | 0.1294 | 0.1525 |

**Table S4. Heterogeneity analysis of disruption percentile across funding types. The sample is divided by funding roles. Standard errors are in parentheses.** *** $p < 0.01$, ** $p < 0.05$, * $p < 0.1$.

| | Gov/Prof | Institution | Publisher | Unfunded |
|---|---|---|---|---|
| $s_{RP}$ | -0.0418*** | -0.0369*** | -0.0317*** | -0.0243*** |
| | -0.0036 | -0.0033 | -0.0045 | -0.0009 |
| $s_{RC}$ | -0.1120*** | -0.1082*** | -0.1078*** | -0.1182*** |
| | -0.0045 | -0.0041 | -0.0059 | -0.0009 |
| $s_{PC}$ | 0.0412*** | 0.0309*** | 0.0399*** | 0.0163*** |
| | -0.0052 | -0.0048 | -0.0068 | -0.001 |
| # of references | 0.0001** | 0.0003*** | -0.0001 | 0.0004*** |
| | -0.0001 | -0.0001 | -0.0001 | 0 |
| Team size | -0.0001 | -0.0001* | -0.0001* | -0.0002*** |
| | 0 | 0 | 0 | 0 |
| # of funders | -0.0036*** | -0.0032*** | -0.0034*** | 0 |
| | -0.0009 | -0.0009 | -0.0011 | 0 |
| Year author most citation percentile | -0.1567*** | -0.1420*** | -0.1141*** | -0.1661*** |
| | -0.0106 | -0.0099 | -0.0139 | -0.0024 |
| const | 0.5739*** | 0.5305*** | 0.5826*** | 0.5870*** |

| | -0.0066 | -0.0064 | -0.0077 | -0.0014 |
|---|---|---|---|---|
| Observations | 28940 | 31801 | 20087 | 428511 |
| Pseudo R-squared | 0.0564 | 0.0563 | 0.0458 | 0.0897 |

**Table S5. Comparison of explanatory power (Pseudo $R^2$) between baseline models and full semantic models across different embedding modalities. The baseline models include only the non-semantic control variables (e.g., number of references, team size, publication year, and author status). The full models integrate the semantic measurements ($s_{RP}$, $s_{PC}$, $s_{RC}$).**

| Dependent Variable | Baseline model | Full model (SPECTER) | Full model (Pure text SciBERT) |
|---|---|---|---|
| Citation percentile | 0.0729 | 0.1493 | 0.1467 |
| Disruption percentile | 0.0278 | 0.0840 | 0.0414 |